\documentclass[a4paper,fleqn,usenatbib]{mnras}

\usepackage[T1]{fontenc}
\usepackage{ae,aecompl}

\usepackage{graphicx}	% Including figure files
\usepackage{amsmath}	% Advanced maths commands
\usepackage{amssymb}	% Extra maths symbols
\usepackage{tabularx,booktabs,tablefootnote}
\usepackage{multirow}
\usepackage{hyperref}
\usepackage{newtxtext,newtxmath}
\newcommand{\eu}[5]{\mbox{$#1\,^#2{\rm #3}^{#4}_{\rm #5}$}}
\newcommand{\vmicro}{$\xi_{\rm t}$}
\newcommand{\kms}{km\,s$^{-1}$}
\newcommand{\te}{$T_{\rm eff}$}
\newcommand{\logg}{$\log{g}$}
\newcommand{\vsini}{$v\sin{i}$}
\newcommand{\vmacro}{$\zeta_{\rm RT}$}

\title[Young CP star BD+30$\degr$549]{BD+30$\degr$549: young helium-weak silicon star in NGC 1333 star-forming region}

\author[I. Potravnov et al.]{
I. Potravnov$^{1}$\thanks{E-mail: ilya.astro@gmail.com (IP)},
L. Mashonkina $^{2}$,
T. Ryabchikova$^{2}$,
\\
$^{1}$Institute of Solar-Terrestrial Physics, Siberian branch of Russian Academy of Sciences, Lermontov Str. 126A, 664033, Irkutsk, Russia\\
$^{2}$Institute of Astronomy of RAS, Pyatnitskaya str., 48, 119017, Moscow, Russia\\}

\date{Accepted XXX. Received YYY; in original form ZZZ}

\pubyear{2023}

\begin{document}
\label{firstpage}
\pagerange{\pageref{firstpage}--\pageref{lastpage}}
\maketitle

\begin{abstract}
We present results of the spectroscopic study of the chemically peculiar star BD+30$\degr$549 which is bona-fide member of young NGC 1333 star forming region. We found that the star possesses negligible rotation and helium-weak spectroscopic pattern with strongly enhanced \ion{Si}{II} and \ion{Si}{III} lines. The fundamental parameters of the star \te=13100~K and $\log (L/L_{\odot})$=2.1 indicate its age of about 2.7 Myr and position on the Hertzsprung-Russell diagram at the end of the Pre-Main Sequence evolutionary track, close to the Zero Age Main Sequence. Abundance analysis reveals the modest deficit of almost all elements with exception of \ion{Si}{}, \ion{Fe}{}, \ion{Ca}{} and \ion{P}{} which are overabundant. We performed the non-LTE calculations for \ion{Si}{II}/\ion{Si}{III}, \ion{Mg}{II} and \ion{Ca}{II} in order to check the influence of departures from LTE on line formation. Non-LTE calculations lead to much better reproduction of individual silicon line profiles, but does not completely remove the abundance discrepancy between \ion{Si}{II} and \ion{Si}{III} lines. We also investigate the effects of possible chemical stratification in BD+30$\degr$549. We suspect that the "\ion{Si}{II}/\ion{Si}{III} anomaly"\, observed in BD+30$\degr$549 spectrum arises under the combined action of the vertical and horizontal abundance gradients and non-LTE effects. We suppose that evolutionary status and phenomena observed in BD+30$\degr$549 indicate that conditions favorable for the selective diffusion and formation of the surface chemical composition peculiarities (slow rotation and stabilization of the atmosphere) can be built up already at the Pre-Main Sequence phase.
\end{abstract}


%%%%%%%%%%%%%%%%%%%%%%%%%%%%%%%%%%%%%%%%%%%%%%%%%%

%%%%%%%%%%%%%%%%% BODY OF PAPER %%%%%%%%%%%%%%%%%%

%%%%%%%%%%%%%%%%%%%%%%%%%%%%%%%%%%%%%%%%%%%%%%%%%%%%%%%%%%%%%%%%%%%%%%%%
\section{Introduction}
%%%%%%%%%%%%%%%%%%%%%%%%%%%%%%%%%%%%%%%%%%%%%%%%%%%%%%%%%%%%%%%%%%%%%%%%

About 15-20$\%$ of stars on the upper Main sequence (MS) represent peculiar spectra which are indicative for the severe anomalies in surface chemical composition \citep[see][for review]{Smith_1996,Romanyuk_2007}. The subgroup of the helium-peculiar stars with the weakened or vice versa enhanced helium lines are considered as an extension of the sequence of magnetic Ap/Bp stars in the region of higher temperatures. The helium-peculiar stars are encountered among the spectral classes $\sim$B3-B8 and many of them also host the large-scale magnetic fields with the intensity $\sim10^3$ G strength. It is believed that the vertical abundance gradients in atmospheres of magnetic chemically peculiar (CP) stars and, hence, inhomogeneities of their surface elemental composition are developed by the atomic diffusion processes \citep{Michaud_1970, Michaud_2015}. It is clear that the diffusion can be sufficiently efficient and leads to the build-up of observable strong abundance inhomogeneities only in absence of the macroscopic mixing caused by convection, turbulence or meridional circulation due to axial rotation of the star.

The exact evolutionary stage when atmospheric stabilization and formation of the vertical abundance gradients occurs is still unknown. The vast majority of CP stars are the MS objects \citep{Kochukhov_2006}. However, there are number of evidence from the evolutionary trend in abundances \citep{Bailey_2014}, frequency of CP stars in clusters of different ages \citep{Netopil_2015} as well as from the theoretical expectations that the surface chemical inhomogeneities could be developed before the star reached the Zero Age MS (ZAMS). The special searches of CP stars were conducted by \citet{Folsom_2012} among Herbig Ae/Be stars which are still contracting Pre-MS (PMS) progenitors of the MS intermediate mass stars. They found several stars with $\lambda$~Boo peculiarity and only one star with weak Ap/Bp pattern - V380 Ori. Recently the weak magnetic field and Ap spectral pattern were discovered in the secondary component of the PMS binary AK Sco \citep{Castelli_2020}. Except these two  PMS objects the present sample of very young CP stars is still very limited and contains few members of young clusters which are believed to settle on the ZAMS very recently \citep[e.g.][]{Bagnulo_2004,Netopil_2014}. The expansion of this sample is essential for better understanding of the timescale and mechanisms which lead to stabilization and formation of the abundance gradients in the atmospheres of early type stars.

One possibility for increasing the sample is based on the data stored in the literature and spectral archives. In discussion on the problem of young CP stars \citet{Herbig_2006} pointed out few poorly investigated stars as possible candidates, including BD+30$\degr$549 which is the subject of the present study. The star BD+30$\degr$549 locates in the northern part of NGC 1333 star forming region and illuminates the eponymous reflection nebula also known as VdB 17 \citep{VdB_1966}. The designation of NGC 1333 star forming region refers to the young stellar cluster associated with complex reflection nebula and dark cloud B255 (or L1450). NGC 1333 lies on the western edge of Perseus molecular cloud at average distance 293$\pm$22 pc \citep{Ortiz_Leon_2018}. It is the most active site of the ongoing low- to intermediate mass star formation within Perseus cloud \citep{Bally_2008,Walawender_2008} with an age of its stellar population ranged from $\sim10^5$ to $5\times10^7$ yr \citet{Aspin_2003}. The high fraction of embedded protostars and disk-bearing stars, Herbig-Haro objects and molecular outflows in this region is consistent with the more recent estimate of the cluster's median age as $t\approx$1 Myr \citep{Luhman_2016}. At this age the stars later than $\sim$B5 should be still PMS or early MS objects.

In early studies BD+30$\degr$549 was classified as B9-8Vp star \citep{Hubble_1922,Racine_1968}. However, no indication of the type of peculiarity and any description of its spectrum was given in these papers. Later BD+30$\degr$549 was often included in samples of Herbig Ae/Be stars, satisfying the fundamental criteria for early spectral type and association with the reflection nebula. Though the presence of emission lines in the spectrum which are indicators of the ongoing mass accretion from the circumstellar disk, and hence definite attribution BD+30$\degr$549 to Herbig Ae/Be group, remains confusing. The index $e$ appeared after the spectral classification of BD+30$\degr$549 in the list of Cep R1 members in \citet{Racine_1968}, but was omitted in his Table II. To the best of our knowledge the only high-resolution spectrum of BD+30$\degr$549 in modern era was obtained by G.~Herbig, but remained unpublished. According to the characteristics given above, the star is a promising candidate for PMS chemically peculiar star. The aim of the present study is to determine the parameters of BD+30$\degr$549 atmosphere, age and evolutionary status, and also to quantitatively characterize its chemical composition.

%%%%%%%%%%%%%%%%%%%%%%%%%%%%%%%%%%%%%%%%%%%%%%%%%%%%%%%%%%%%%%%%%%%%%%%%
\section{Observations and data reduction}
%%%%%%%%%%%%%%%%%%%%%%%%%%%%%%%%%%%%%%%%%%%%%%%%%%%%%%%%%%%%%%%%%%%%%%%%
To our analysis we used the high resolution optical spectrum of BD+30$\degr$549 retrieved from Keck Observatory archive\footnote{\url{https://www2.keck.hawaii.edu/koa/public/koa.php}}. The star was observed with the Keck I telescope and High Resolution Echelle Spectrometer (HIRES) on 2 Feb. 2000 (PI: G.Herbig). The observations were carried out with the 0.86$\arcsec$ projected slit width that resulted in nominal resolving power about $R\approx48000$. The $\lambda\lambda$4240-6710~\AA\, range was covered by spectrogram with some interorder gaps in the red. The signal-to-noise ratio was estimated in the region near $H\alpha$ as S/N$\approx$130 (per pixel). The raw data were processed with the MAKEE\footnote{\url{https://sites.astro.caltech.edu/~tb/makee/}} (\textsc{MAuna Kea Echelle Extraction}) pipeline written by T. Barlow. The data processing workflow included the bias subtraction and flatfielding of the science frames as well as wavelength calibration using reference spectrum of Th-Ar hollow cathode lamp. After this calibration the 1D spectrum extraction was performed. The accuracy of the localization of spectral orders in the 2D image, boundaries of object and background extraction were manually controlled.

A well-known difficulty in processing the echelle spectra of the early type stars is the normalization of the orders containing broad hydrogen lines to the continuum level. We used the method of interpolating a continuum from the neighboring orders traced with low-degree cubic spline. Thanks to the relatively smoothness of HIRES blaze function application of this method resulted in reasonably accurate normalization of the wings of hydrogen lines. The only exception was the very broad and weak depression in the red wing of H$\beta$ centered roughly at 4880~\AA\, which is presumably of interstellar origin. Indeed this feature coincide in wavelength with the very broad and shallow diffuse interstellar band (DIB) at 4882~\AA\, \citep{Herbig_1995,Galazutdinov_2020}. In any case, this region was excluded from the further spectroscopic analysis.

%%%%%%%%%%%%%%%%%%%%%%%%%%%%%%%%%%%%%%%%%%%%%%%%%%%%%%%%%%%%%%%%%%%%%%%%
\section{Results}
%%%%%%%%%%%%%%%%%%%%%%%%%%%%%%%%%%%%%%%%%%%%%%%%%%%%%%%%%%%%%%%%%%%%%%%%

\subsection{Stellar parameters, distance, extinction}

The absorption line-spectrum of BD+30$\degr$549 generally corresponds to the mid- to late B photosphere. No hydrogen or metallic lines emissions indicative for the accretion activity were detected. Hence BD+30$\degr$549 cannot be classified as typical Herbig Ae/Be star at least on the base of the spectrogram at our disposal. Its comparison with the spectra\footnote{High resolution ($R$=42000) spectra were retrieved from the ELODIE archive \citep{Moultaka_2004}} of two late-B MK standards \citep{Gray_Corbally_2009}, namely HR1029 (sp:B7V) and 134 Tau (sp:B9IV) revealed the striking absence of the \ion{He}{I} lines, even the strongest ones at 4471 and 5876~\AA\,. The strong \ion{Mg}{II} 4481~\AA\, line visible in spectrum of of BD+30$\degr$549 seems to be somewhat shallow comparing to that in 134 Tau spectrum. At the same time the \ion{Si}{II} lines at 4621, 5041, 5055, 5466, 6347, 6371~\AA\, are abnormally enhanced. The \ion{Si}{III} triplet at 4552, 4567, 4574~\AA\, which is almost absent in the spectra of standards clearly seen in the spectrum of BD+30$\degr$549. The notable sharpness of the absorption lines of different metals indicates very slow rotation. Qualitatively it is possible to classify BD+30$\degr$549 as helium-weak Bp star with silicon peculiarity.

Rich interstellar spectrum which contains both atomic and molecular features as well as numerous DIBs is superimposed on the photospheric ones. Measurements of the radial velocity of the CH at 4300~\AA\,, \ion{Na}{I} D interstellar features and few most symmetric DIBs including those at 5780, 5797, 6380, 6614~\AA\, provide average value $RV_h\approx$+14~\kms\ which is close to the heliocentric radial velocity $RV_h$=+16.45~\kms of BD+30$\degr$549 measured by cross-correlation of the photospheric absorption lines with the template spectrum in several spectral windows.

The \textit{Gaia} EDR3 \citep{GAIA_2021} provides the parallax $\pi=3.48\pm0.0237$ for BD+30$\degr$549 with Renormalised Unit Weight Error parameter $RUWE=1.14$. This value of RUWE is well below the recommended 1.4 upper threshold\footnote{\url{https://dms.cosmos.esa.int/COSMOS/doc_fetch.php?id=3757412}} and indicates the "good behaved"\, astrometric solution. The parallax of the star can be converted into distance $D=287\pm2$ pc which is in the perfect agreement with the average distance to the NGC 1333 cluster. Both the spatial position and coincidence of the radial velocity of the star with the mean velocity of molecular material in NGC 1333 region $RV_h\approx+15.7$ km$\cdot$s$^{-1}$ determined from the ${}^{13}$CO (3-2) and C${}^{18}$O (3-2) transitions \citep{Curtis_2010} confirms that location of BD+30$\degr$549 inside the reflection nebula is not due to the projection effect. The star is bona fide member of NGC 1333 star forming region and still intimately connected with its parental molecular material. 

Embeddedness inside the reflection nebula implies the non-negligible effects of interstellar reddening which affect the observed photometry of BD+30$\degr$549. The variations of the ratio of total to selective absorption $R_V$ within NGC 1333 was investigated by \citet{Cernis_1990} who found that in direction toward BD+30$\degr$549 it deviates significantly from the interstellar one and equals to $R_V=4.7\pm0.1$. In order to determine the value of the interstellar extinction $A_V$ we used the color excess $E(B-V)=+0.59^m$ calculated from the observed color index $(B-V)$=+0.48$^m$ \citep{Henden_2016} and the intrinsic colors interpolated from the \citet{Pecaut_Mamajek_2013} tables for our best-fit effective temperature \te=13100~K (Sect. \ref{sect:atm_param}). As a result we obtained visual extinction $A_V=2.8^m$. The observed spectrophotometry of BD+30$\degr$549 was dereddened using this value and \citet{Fitzpatrick_2019} extinction curve. The bolometric absolute magnitude $M_\mathrm{bol}^{*}=-0.5^m$ was found for BD+30$\degr$549 adopting 287 pc distance and using this extinction value as well as the bolometric correction $BC_V=-0.9^m$ \citep{Pecaut_Mamajek_2013}. Finally, the absolute magnitude was converted to luminosity: $\log (L_*/L_{\odot})= 0.4(M_\mathrm{bol}^\odot-M_\mathrm{bol}^{*})=2.1\pm0.1$. 

\begin{figure}
	\includegraphics[width=1.0\linewidth]{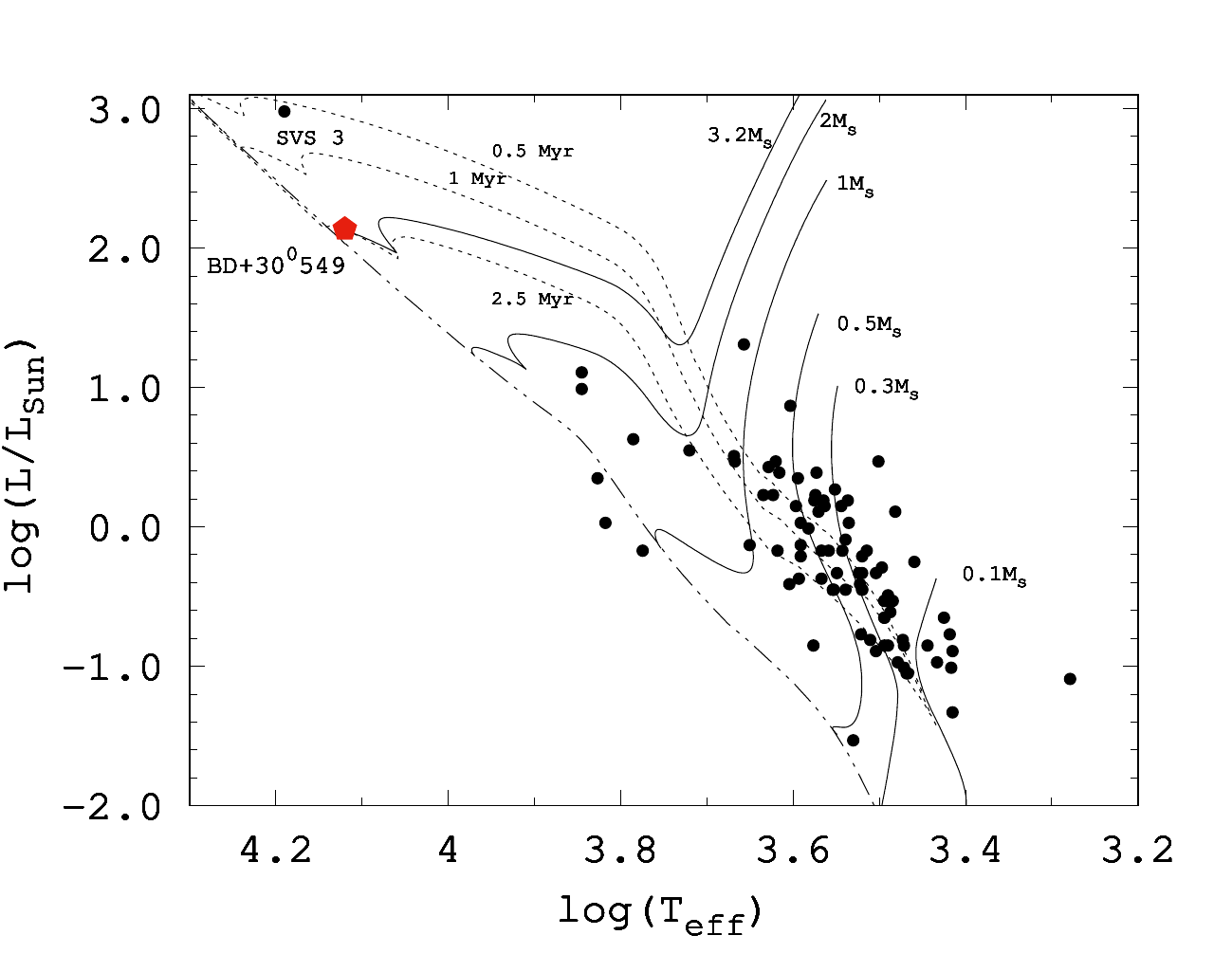}
    \caption{HR diagram with position of BD+30$\degr$549 marked by diamond symbol. The sample of low-mass members of the NGC 1333 cluster from \citet{Foster_2015} is shown by small dots as well as the most massive cluster member $\sim$B5 star SVS3. Theoretical evolutionary tracks and isochrones from the PARSEC grid \citep{Bressan_2012} are represented by solid and dotted curves respectively and labeled. ZAMS is shown by dash-dotted curve.    }
    \label{fig:1}
\end{figure}

With the obtained values of luminosity and effective temperature the star was placed on HR diagram (Fig. \ref{fig:1}). Comparison with the evolutionary tracks and isochrones from the PARSEC model grid \citep{Bressan_2012} calculated for solar metallicity $Z=0.017$ revealed that BD+30$\degr$549 lies near the end of the $3.2M_{\odot}$ PMS track, close to ZAMS. The stellar radius inferred from the model track is $R=2.2R_{\odot}$. The age of the star was determined from the closest theoretical isochrone as $t\approx$2.7 Myr.

%%%%%%%%%%%%%%%%%%%%%%%%%%%%%%%%%%%%%%%%%%%%%%%%%%%%%%%%%%%%%%%%%%%%%%%%%%%%%%%
%%%%%%%%%%%%%%%%%%%%%%%%%%%%%%%%%%%%%%%%%%%%%%%%%%%%%%%%%%%%%%%%%%%%%%%%%%%%%%%

\subsection{Spectrophotometric properties and variability of BD+30$\degr$549}\label{sect:phot}
\subsubsection{Spectral energy distribution}
\begin{figure}
	\includegraphics[width=1.0\linewidth]{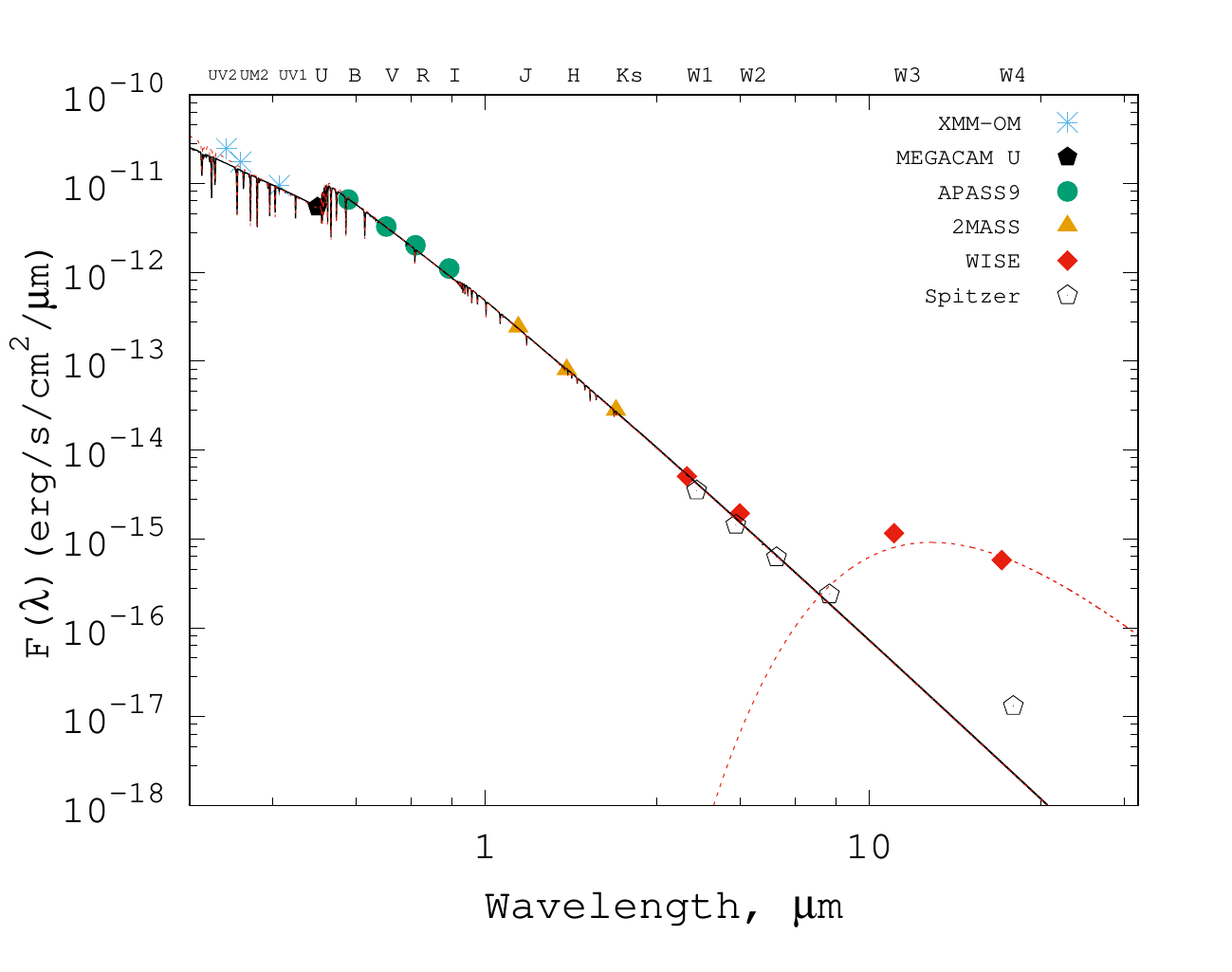}
    \caption{Dereddened SED of BD+30$\degr$549 from UV to mid-IR. Observed spectrophotometry from different sources are shown with different types of dots, explained in the legend. Solid black curve represents theoretical SED calculated with final set of atmospheric parameters, namely for \te=13100~K, \logg=4.2 and helium-weak chemical composition. The emergent flux calculated with the same parameters but for stratified model is shown by dotted curve. Dashed curve (red in the electronic version) represents the Plank function for \te=200~K. }
    \label{fig:2}
\end{figure}

The spectral energy distribution (SED) of BD+30$\degr$549 was constructed in the 0.2-24 $\micron$ wavelength range (from ultraviolet (UV) to mid-infrared (mid-IR)) using the photometric data from XMM-OM survey \citep{Page_2012}, APASS-9 \citep{Henden_2016}, 2MASS \citep{Skrutskie_2006}, WISE \citep{Wright_2010} catalogs and Spitzer photometry \citep{Evans_2003}. Comparison of the photospheric flux calculated with our final set of atmospheric parameters (Sect. \ref{sect:atm_param}) and dilluted for the given distance $D=287$ pc with dereddened observations is shown in Fig. \ref{fig:2}. Previously, BD+30$\degr$549 was classified as $J$-type source that implies the presence of near-IR (NIR) excess due to thermal radiation of circumstellar dust \citep{Guzman_2021}. With a newly adopted \te\ and specific reddening our SED fitting resulted in excellent match between the theoretical flux and observed photometry in the optical and NIR regions within 0.23-8 $\micron$ range. However, the observed SED definitely shows the UV excess in the XMM-OM $UVW2$, $UVM2$ bands as well as mid-IR excess at $\lambda\gtrsim$12 $\micron$ in the $W3-4$ and MPIS 24 $\micron$ bands. While the UV excess could be caused by the anomalous temperature structure of the BD+30$\degr$549 atmosphere (Sect. \ref{sect:si_impact}), the mid-IR excess is reasonably well fitted by the $\sim$200~K blackbody radiation, as one can see from Fig. \ref{fig:2}. The peak temperature of the excess is about an order of magnitude higher than the typical $\sim10^1$~K dust temperature in the reflection nebulae \citep[e.g.][]{Gibson_2003}. Thus the observed mid-IR excess traces the radiation of the warm dust located closer to the star in the circumstellar disk surrounding BD+30$\degr$549. At the same time, the lack of NIR excess in the \textit{JHK} bands and absence of the spectroscopic signatures of accretion indicates that gas and dust are depleted in the immediate vicinity of the star. Such a radial distribution of the circumstellar material is characteristic for the rather evolved disks: transitional or debris ones \citep{Williams_2011}

A noteworthy detail in the mid-IR part of the SED of BD+30$\degr$549 is the step difference in flux between two nearby $W4$ and MPIS 24 $\micron$ points. The differences of about order of magnitude exceed the errors of both WISE and Spitzer photometry, thus most probably it reflects the real variability of $\sim24$ $\micron$ flux. We have inspected the WISE multiepoch photometry of BD+30$\degr$549 obtained on a few dates within a 6-month interval in 2010, but were not able to detect any variability exceeding 0.2$^m$. However, the Spitzer data we used were obtained in the course of original Cores-to-Disks (c2d) Legacy program \citep{Evans_2003}. Thus, variations in 24 $\micron$ flux occurred on a decadal timescale.
A possible explanation of this mid-IR variability is the appearance of a large amount of dust in the disk at some point between the Spitzer and WISE observations. As simulations show \citep{Kenyon_Bromley_2005}, the amount of dust which can significantly affects the observed excess at 24 $\micron$ could arise from the collision of $\sim$1000-km size bodies: planetesimals or planetary embryos during the late stages of planet formation. Indeed such catastrophic events resulting in temporary rise of mid-IR excess are sometimes observed in young stars with planet-forming disks \citep[e.g.][]{Melis_2012,Su_2022}.

%%%%%%%%%%%%%%%%%%%%%%%%%%%%%%%%%%%%%%%%%%%%%%%%%%%%%%%%%%%%%%%%%%%%%%%%%%%%%%%
%%%%%%%%%%%%%%%%%%%%%%%%%%%%%%%%%%%%%%%%%%%%%%%%%%%%%%%%%%%%%%%%%%%%%%%%%%%%%%%

%%%%%%%%%%%%%%%%%%%%%%%%%%%%%%%%%%%%%%%%%%%%%%%%%%%%%%%%%%%%%%%%%%%%%%%%%%%%%%%
\begin{figure}
	\includegraphics[width=1.0\linewidth]{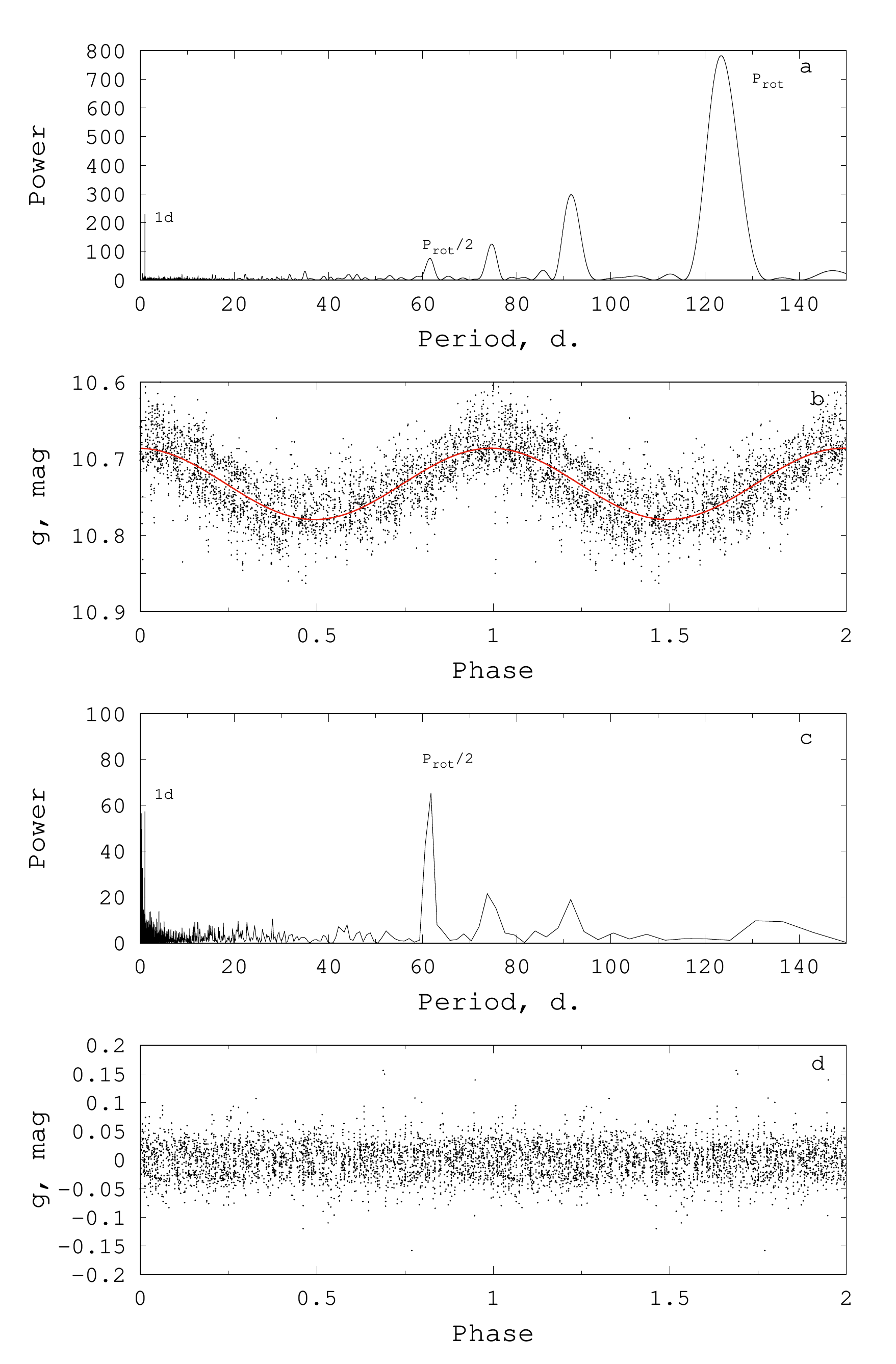}
    \caption{Upper panel (a): power spectrum for ASAS-SN archival photometry of BD+30$\degr$549. Panel (b): phase curve folded with 123$^d$.3 period. The sinusoidal fit is shown by curve. Panel (c): power spectrum after subtraction the sinusoidal fit corresponding to 123$^d$.3 period. Lower panel (d): deviation from the mean curve after subtraction the main rotational signature and its first harmonic as well as correction for the diurnal variations.}
    \label{fig:3}
\end{figure}
%%%%%%%%%%%%%%%%%%%%%%%%%%%%%%%%%%%%%%%%%%%%%%%%%%%%%%%%%%%%%%%%%%%%%%%%%%%%%%%
\subsubsection{Photometric variability}\label{sect:phot_var}
In order to investigate the long-term photometric behavior of BD+30$\degr$549, we retrieved its light curve from the database of All-Sky Automated Survey for Supernovae (ASAS-SN) \citep{Kochanek_2017}. Observations in Sloan $g$ filter covering six observational seasons from 2017 to 2022 were analyzed. Even at the first glance, the periodic changes in the stellar brightness with an $\Delta V\approx0.2^m$ amplitude were manifested in the lightcurve. The frequency analysis yielded several peaks on the Lomb-Scargle periodogram, the most powerful of which correspond to $\approx$123$^d$.3 period (Fig. \ref{fig:3}, panel a). The phase curve folded using the ephemeris $JD(max.light)=2458101.677+123^d.277\cdot E$~ is also shown in Fig. \ref{fig:3} (panel b). One can see the smooth quasi-sinusoidal light changes with the prolonged minimum and somewhat sharper maximum. 

Similar variability is common within the Ap/Bp stars with elemental spots on their surface. For example, well studied Ap-Si star CU Vir shows quasi-sinusoidal light changes with an amplitude up to $0.2^m$ in $U$ filter \citep{Pyper_1998}. The brightness maximum of CU Vir coincides with the maximum of absorption in the \ion{Si}{II} lines. If we assume that the photometric variability of BD+30$\degr$549 is also rotationally-modulated and caused by the chemical spots with an altered temperature structure, the date of spectral observation falling on the phase $\phi=0.94$ should also corresponds to the line absorption maximum. It is essential to obtain additional spectroscopic observations near the phase $\phi \approx0.5$ (corresponding to the minimum light) to test this hypothesis. 

Except the main peak at the stellar rotational period $P_{rot}$ we identified its first harmonic $P_{rot}/2$ and few aliases at 73$^d$.7, 91$^d$.5 and 1$^d$. After subtraction rotationally-modulated signal corresponding to the main peak, aliases at 73$^d$.7 and 91$^d$.5 became insignificant, however periodogram still contained meaningful peaks corresponding to the first harmonic of rotational period $P_{rot}/2$ and diurnal alias (Fig. \ref{fig:3}, panel c). Further subtraction of the latter two signals from the data resulted in the 3-$\sigma$ scatter of residuals around the mean of about $\sim0.09^m$ (Fig. \ref{fig:3}, panel d). This is typical accuracy of the ground-based patrol photometry. However, it is worth noting that BD+30$\degr$549 lies in the temperature domain of the Slowly Pulsating B (SPB) stars with periods typically ranged from $\sim0.5^d$ to 4$^d$ \citep{Waelkens_1991,Kurtz_2022}. Whether BD+30$\degr$549 possesses such type of variability or not could be clarified with high-precision space photometry.

\subsection{Atmospheric parameters and average abundances}\label{sect:atm_param}

%%%%%%%%%%%%%%%%%%%%%%%%%%%%%%%%%%%%%%%%%%%%%%%%%%%%%%%%%%%%%%%%%%%%%%%%%%%%%%%
\begin{figure*}
	\includegraphics[width=1.0\linewidth]{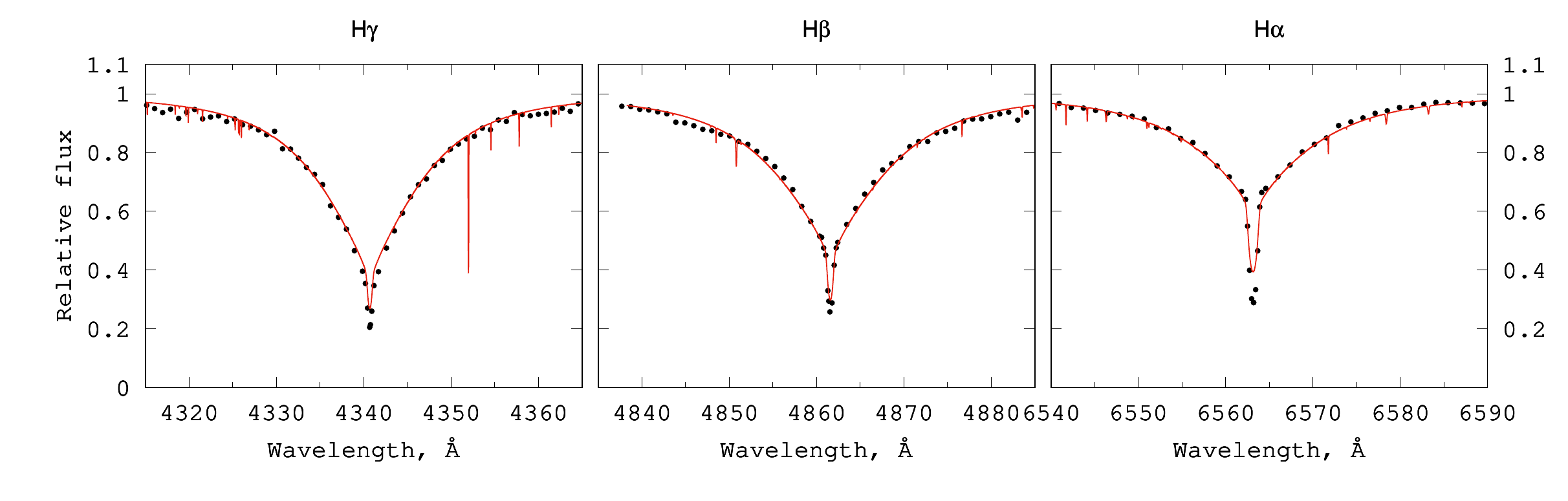}
    \caption{Balmer line profiles in BD+30$\degr$549 (dotted curve) compared with the best fit synthetic spectrum (continuous curve). }
    \label{fig:4}
\end{figure*}
%%%%%%%%%%%%%%%%%%%%%%%%%%%%%%%%%%%%%%%%%%%%%%%%%%%%%%%%%%%%%%%%%%%%%%%%%%%%%%%

The atmospheric parameters of BD+30$\degr$549 were determined using the self-consistent iterative procedure based on spectral synthesis technique. As an initial guess, we used the effective temperature and surface gravity of a B8 star from the MK system calibrations \citep{Gray_Corbally_2009,Kaler_1982}. In the next step, the grid of the helium-weak atmospheric models with hydrogen abundance $N_H/N_{tot}=0.99$ was calculated using the \textsc{LLmodels} code \citep{Shulyak_2004}, which accurately accounts the influence of anomalies in chemical composition on the opacity distribution in the atmosphere. Hereafter, we use the models with the temperature structure calculated for the homogeneous abundance distribution as the best approximation to date (see Sect. \ref{sect:si_impact} for discussion). Using these model atmospheres and atomic data extracted from the VALD3 database \citep{Piskunov_1995,Ryabchikova_2015,Pakhomov_2019} the synthetic spectra were calculated with the \textsc{Synth3} code \citep{Kochukhov_2007}. The \te, \logg~ parameters had been refining by the comparison of the synthetic spectrum with the observed one in the region of the wings of hydrogen H$\alpha$, H$\beta$ and H$\gamma$ lines. Since in late-B stars the hydrogen lines are both temperature- and gravity-sensitive we used the observed SED as the independent temperature constraint. The observed spectrophotomery was dereddened using \citet{Fitzpatrick_2019} extinction curve adopting $R_V=4.7$. The dereddened SED was compared with the synthetic fluxes computed on each iteration with \textsc{LLmodels} code for a given model atmosphere. The theoretical fluxes were dilluted assuming the 287 pc distance and the stellar radius $R=2.2R_{\odot}$. The microturbulent velocity \vmicro~ was determined by the classical method of minimizing the slope of the dependence of the elemental abundance on the equivalent widths. We used a set of unblended \ion{Fe}{II}/\ion{}{III} lines further employed for stratification analysis (see Sect. \ref{sect:strat} and Tab. \ref{tab3}). The calculations were performed with a version of the Kurucz's \textsc{Width} code modified by V. Tsymbal (priv.com.). The procedure of spectrum and SED fitting was repeated in a few iterations. The uncertainties of derived parameters were estimated from the dispersion of few last iterations around the best-fit solution.

As a result the following set of parameters was determined: the effective temperature \te=13100$\pm$100~K, the surface gravity \logg=4.2$\pm0.1$, the microturbulent velocity \vmicro=0~\kms. The stellar parameters are also summarised in Table \ref{tab1}. The comparison of the observed and synthetic Balmer lines profiles calculated with this final set of parameters is shown in Fig. \ref{fig:4}. One can see the reasonable agreement between synthetic spectrum and observations for H$\beta$ and H$\gamma$ lines. The prominent discrepancy between the
observations and synthetic spectrum in the core of the H$\alpha$ line is caused by unaccounted non-LTE effects, while a little mismatch in the red wing could be due to the spectrum reduction faults. At the same time, the theoretical SED reproduces well the magnitude of the Balmer discontinuity and the slope of the Paschen continuum (Fig. \ref{fig:2}), that confirms the reliability of our effective temperature determination.

The macroscopic broadening parameters: rotational broadening \vsini\ and macroturbulence \vmacro\ were found by fitting the metallic lines profiles in several spectral windows (e.g. 4590-4650\,\AA\,, 5045-5115\,\AA\,) with the \textsc{BinMag6} tool \citep{Kochukhov_2018b}. In agreement with the visual impression of the lines sharpness, neither rotational nor macroturbulence broadening was detected. Given the spectral resolution and instrumental profile width we can set the upper limit for rotational velocity as \vsini$\lesssim$1-2~\kms\,. 

Despite the sharpness of absorption lines and the evidence for highly stabilized atmosphere no sign of Zeeman splitting or sufficient magnetic intensification was detected in the magnetically-sensitive spectral lines like \ion{Fe}{II} $\lambda\lambda$4303, 4385, 4520, 6149~\AA\,. Within the measurement error we also were unable to detect the differential magnetic intensification in the \ion{Fe}{II} 6147/6149\,\AA\, pair of lines which often is considered as an indication of the magnetic field presence \citep{Mathys_1992,Kochukhov_2013}. These lines have identical intensity in normal stars without magnetic fields but possess a different Zeeman spitting that produces a difference in the observed equivalent widths when the magnetic field is presented.  
We estimate the normalized equivalent width difference of these lines as $\Delta W_{\lambda}/\overline{W_{\lambda}}\lesssim0.03$. Unfortunately, there is no accurate experimental transition probabilities for these lines. Non-magnetic spectrum synthesis results in  $\Delta W_{\lambda}/\overline{W_{\lambda}}$=0.026 with the theoretical transition probabilities from \citet{Raassen_1998} while $\Delta W_{\lambda}/\overline{W_{\lambda}}$=0.003 with the theoretical transition probabilities from Kurucz's 2013 line list\footnote{\url{http://kurucz.harvard.edu/atoms/2601/gfemq2601.pos}}. Magnetic spectrum syntesis of this pair with the help of SYNMAST code  \citep{Kochukhov_2007} showed that a global magnetic field $\langle B\rangle >$1~kG produced too wide synthetic line profiles. Therefore, we set an upper limit on the strength of the mean magnetic field as $\langle B\rangle \lesssim$1~kG. Although based on single-epoch spectroscopy, we cannot rule out the possible existence of a stronger magnetic field in BD+30$\degr$549, which could be detected in another rotational phase, or, more confidently, with the spectropolarimetric observations. \\

\begin{table}
\caption{Parameters of BD+30$\degr$549}
\label{tab1}
\begin{tabular}{lc}
\hline
Parameter & Value \\
\hline
\hline
\te & 13100$\pm$100 K \\
\logg  & 4.2$\pm$0.1 dex \\
\vmicro  & 0.0$\pm$0.2~\kms \\
\vmacro  & 0.0$\pm$1.5~\kms \\
\vsini    & <2.0~\kms \\
$\langle B\rangle$& $\lesssim 1$ kGs\\
$A_V$        & 2.8$^m$\\
$\log (L/L_{\odot})$ & 2.13$\pm$0.05\\
$M/M_{\odot}$       & 3.2\\
$R/R_{\odot}$       & 2.2\\
Age           & $\approx$2.7 Myr\\
\hline
\end{tabular}
\end{table}    

Line identification in BD+30$\degr$549 spectrum was performed using a synthetic spectrum calculated in the entire range $\lambda\lambda$4240-6710\,\AA\, covered by the spectrogram at our disposal, we also used the line lists from \citet{Fossati_2009} for guidance. Due to relatively high effective temperature most of the elements in BD+30$\degr$549 spectrum are presented by the moderate number of lines of the first ions. Only iron and silicon possess lines originated from two different ionization stages. The negligible rotational broadening simplified the task of lines selection. We were able to select a reasonable number of lines for abundance analysis, although for some light elements e.g. \ion{C}{} and \ion{O}{} the sampling was incomplete due to the limited spectral coverage and decrease of S/N ratio in the blue spectral region. The atomic data for abundance analysis were taken from the VALD3 database. For the light elements \ion{Mg}{} and \ion{Si}{} the oscillator strengths $gf$, excitation energies and damping constants retrieved from the VALD were checked against critically selected data from \citet{Mashonkina_2018,Mashonkina_2020} and if necessary replaced by the latter values. We compiled a list of \ion{Fe}{II}/\ion{}{III} with excitation energies ranged from 2.8 to 18.2 eV for the abundance determination and subsequent stratification analysis (see Tab. \ref{tab3}). Hyperfine splitting was taken into account for \ion{Al}{II} and \ion{Ti}{II} using the facilities of VALD3 database \citep{Pakhomov_2019}.

Abundances, $\log (A)_X = \log(N_{X}/N_{H})$, were determined under LTE assumption as the mean of measurements of several lines of a given element $X$. Individual abundances were deduced from the fitting of observed line profiles by synthetic one with \textsc{BinMag6} tool. The error in abundance determinations was estimated as a standard deviation of individual measurement from the mean.

%%%%%%%%%%%%%%%%%%%%%%%%%%%%%%%%%%%%%%%%%%%%%%%%%%%%%%%%%%%%%%%%%%%%%%%%%%%%%%%
\begin{figure}
	\includegraphics[width=1.0\linewidth]{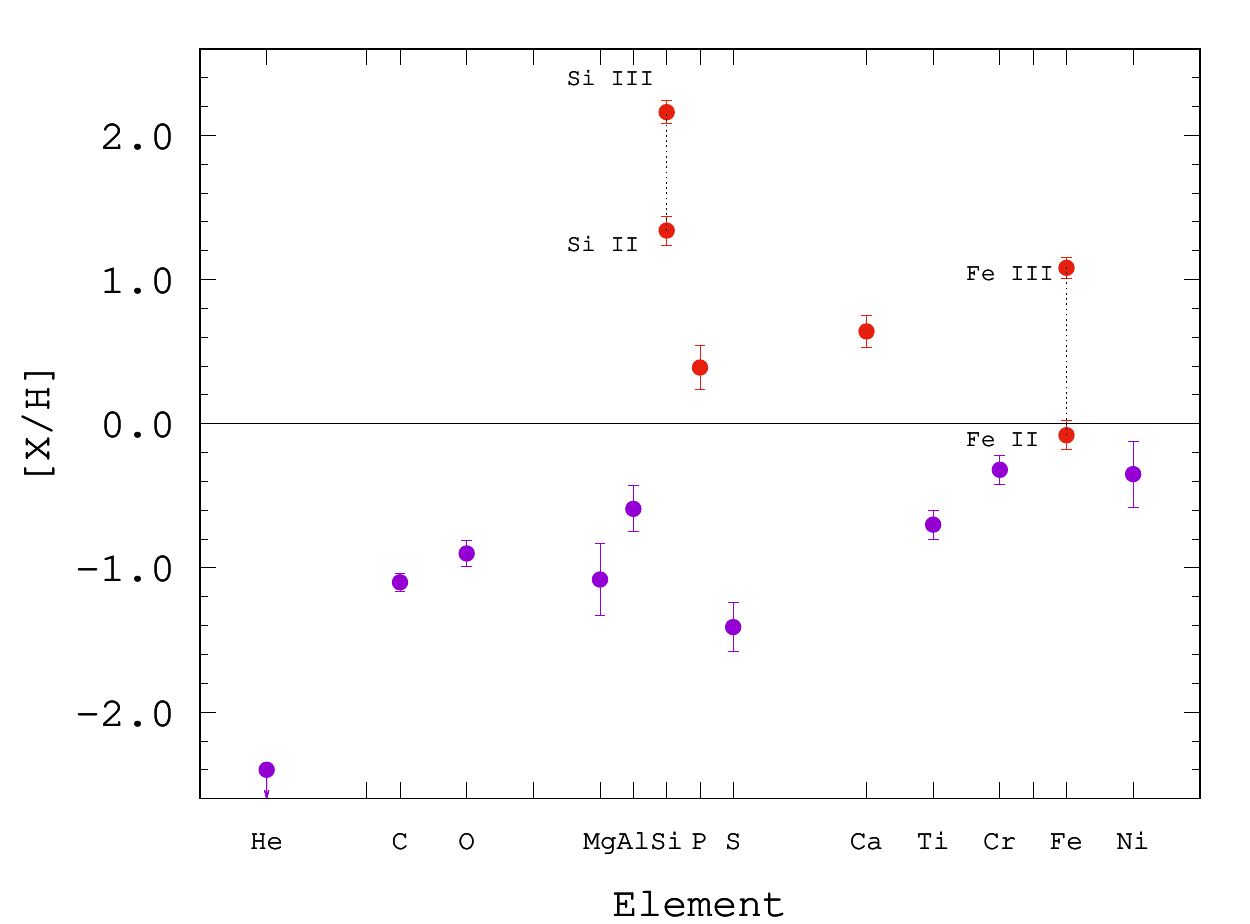}
    \caption{LTE abundances relative to the Sun in BD+30$\degr$549. For \ion{Si}{} and \ion{Fe}{} abundances deduced from the lines of the first and second ions are presented (connected by thin dashed lines). The upper limit for \ion{He}{} abundance is marked.}
    \label{fig:5}
\end{figure}
%%%%%%%%%%%%%%%%%%%%%%%%%%%%%%%%%%%%%%%%%%%%%%%%%%%%%%%%%%%%%%%%%%%%%%%%%%%%%%%

We derived LTE abundances of 13 elements. The results of the abundance analysis are summarised in Table \ref{tab2} and are also shown in Fig. \ref{fig:5}. In the latter plot we compare the derived abundances with the solar ones as [X/H]=$\log(N_{X}/N_{H}) - \log(N_{X}/N_{H})_{\odot}$. The reference solar abundances were taken from \citet{Scott_2015a,Scott_2015b} for elements from \ion{Na}{} to \ion{Ni}{} while the rest were adopted from \citet{Asplund_2009}. Most of the studied elements in BD+30$\degr$549 show depletion up to $\approx1$~dex with respect to the solar atmosphere. We were unable to detect any traces of absorption due to \ion{He}{I} at 4471, 5015, 5876~\AA\,  and hence put the upper limit for helium abundance as $\log (A)_{\ion{He}{}}\leq -3.5$ dex, which is 2.4 dex lower than the solar value. 
Contrary, \ion{Si}{}, \ion{Ca}{}, \ion{P}{} and \ion{Fe}{} display the overabundance. For silicon and iron, we found substantial abundance differences of 0.8~dex and 1.2~dex, respectively, between the lines of the first and second ions. Also for the strongest lines of \ion{Si}{II} (5055/5056, 6347, 6371~\AA) and \ion{Mg}{II} 4481~\AA, their wings and cores cannot be fitted with a single value of abundance. In such cases, the element abundance that fits the entire profile in the best way was adopted as a final value. The found discrepancies inspired us to check an influence of the departures from LTE on line formation for \ion{Si}{ii-iii}, \ion{Mg}{ii}, and \ion{Ca}{ii} and to determine the non-local thermodynamic equilibrium (non-LTE) abundances (Sect.~\ref{sect:nlte}).

Despite the mild phosphorus overabundance in BD+30$\degr$549 atmosphere, we were unable to detect any gallium lines which could be also enhanced in spectra of the phosphorus-gallium subgroup of the helium-weak stars. We also did not find any strontium lines, because the strongest of them, \ion{Sr}{II} 4077\AA\ and 4215~\AA, were not covered by our spectrogram.  

\begin{table}
\caption{Average LTE abundances in BD+30$\degr$549}
\label{tab2}
\begin{tabular}{lccc}
\hline
Ion & $N_{lines}$ & $\log (N_{X}/N_{H})$ & $\log (N_{X}/N_{H})_{\odot}$  \\
\hline
\hline
\ion{He}{I} & ... & $\leq-3.5$ & -1.07 \\
\ion{C}{II} & 2 & -4.67$\pm0.06$ & -3.57 \\
\ion{O}{I} & 3 & -4.21$\pm0.09$ & -3.31 \\
\ion{Mg}{II} & 6 & -5.48$\pm0.25$ & -4.41 \\
\ion{Al}{II} & 4 & -6.16$\pm0.16$ & -5.57 \\
\ion{Si}{II} & 7 & -3.15$\pm0.05$ & -4.49 \\
\ion{Si}{III} & 3 & -2.33$\pm0.16$ & -4.49 \\
\ion{P}{II} & 6 & -6.20$\pm0.15$ & -6.59 \\
\ion{S}{II} & 6 & -6.29$\pm0.17$ & -4.88 \\
\ion{Ca}{II} & 5 & -5.04$\pm0.11$ & -5.68 \\
\ion{Ti}{II} & 6 & -7.77$\pm0.1$ & -7.07 \\
\ion{Cr}{II} & 8 & -6.70$\pm0.1$ & -6.38 \\
\ion{Fe}{II} & 20 & -4.61$\pm0.1$ & -4.53 \\
\ion{Fe}{III} & 7 & -3.44$\pm0.07$ & -4.53 \\
\ion{Ni}{II} & 5 & -6.15$\pm0.23$ & -5.80 \\

\hline
\end{tabular}
\end{table}

\subsection{Non-LTE effects on lines of \ion{Si}{II}--\ion{}{III}, \ion{Ca}{II}, and \ion{Mg}{II}}\label{sect:nlte}

\begin{figure}
\includegraphics[width=\columnwidth]{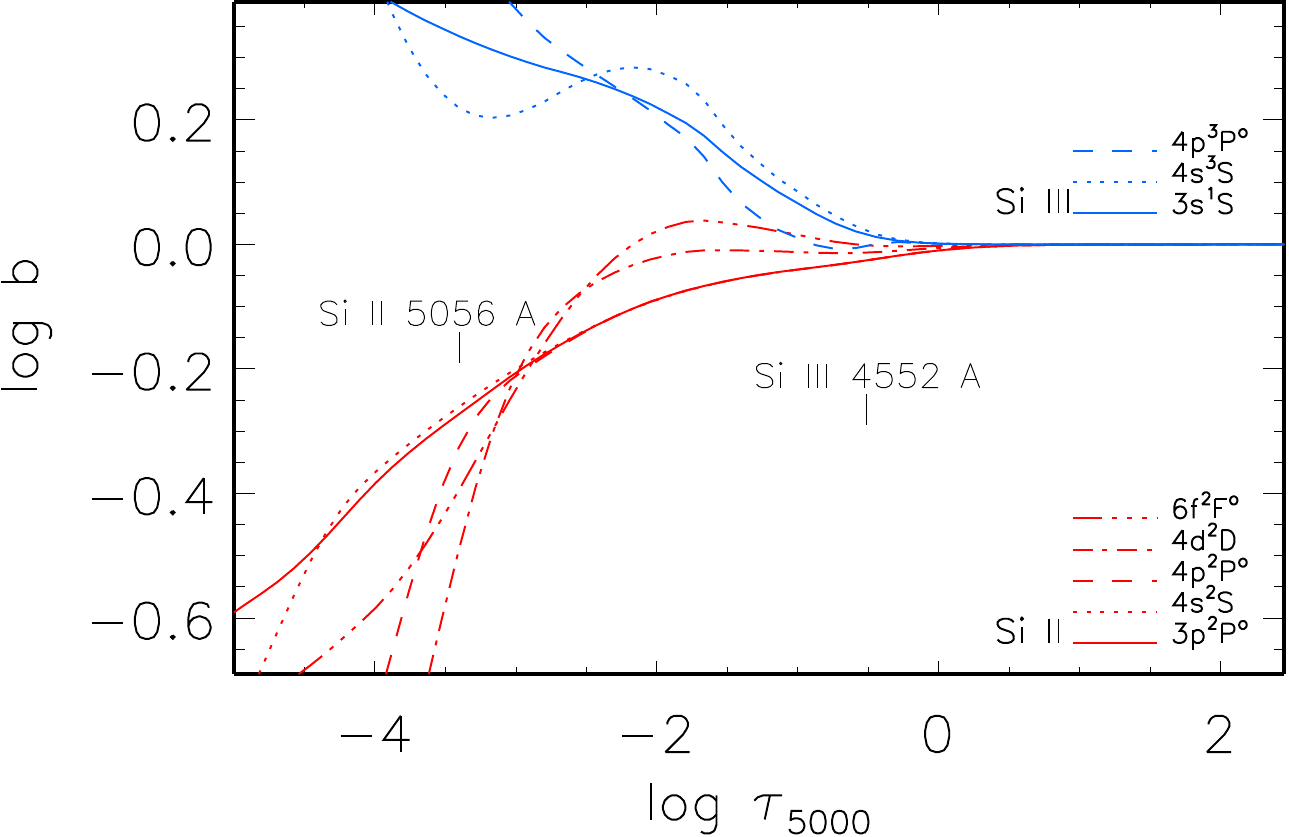}
    \caption{Departure coefficients, $b$, for the selected levels of \ion{Si}{II} (red curves) and \ion{Si}{III} (blue curves) as a function of $\log \tau_{5000}$ in the model atmosphere of BD+30$\degr$549. Tick marks indicate the locations of line center optical depth unity for \ion{Si}{II} 5056\,\AA\ and \ion{Si}{III} 4552\,\AA. Here, we used log$N_{Si}/N_{H} = -2.77$.}
    \label{fig:6}
\end{figure}

Lines of \ion{Si}{II} and \ion{Si}{III} in BD+30$\degr$549 reveal the largest deviations from the classical line-formation scenario that is based on the assumptions of LTE and a chemical homogeneity of the atmosphere. In this section, we check an influence of the departures from LTE on the statistical equilibrium (SE) of silicon and the element abundances derived from different Si lines. The non-LTE calculations were performed with the model atom treated by \citet{Mashonkina_2020}. It includes levels of the first three ionization stages (\ion{Si}{I}, \ion{Si}{II}, and \ion{Si}{III}) and the ground state of \ion{Si}{IV} and implements the most up-to-date atomic data on transition probabilities, photoionization cross-sections, and electron-impact excitation rates. One of the stars studied by \citet{Mashonkina_2020}, namely, HD~17081 ($\pi$~Cet), has atmospheric parameters ($T_{\rm eff}$ = 12800~K and log~$g$ = 3.75) close to that of our star. The LTE analysis of $\pi$~Cet found an abundance difference of 0.23~dex between the two ionization stages, \ion{Si}{II} and \ion{Si}{III}, while consistent within 0.03~dex abundances were obtained in the non-LTE calculations. Compared with $\pi$~Cet, BD+30$\degr$549 reveals substantially larger LTE abundance discrepancies between \ion{Si}{II} and \ion{Si}{III} and high Si abundance, which exceeds the solar one by more than 1.3~dex. In the atmosphere enhanced with silicon, the Si line-formation depths shift to the upper  atmospheric layers that can result in the stronger non-LTE effects compared with those for $\pi$~Cet.

\begin{figure}           
\includegraphics[width=\columnwidth]{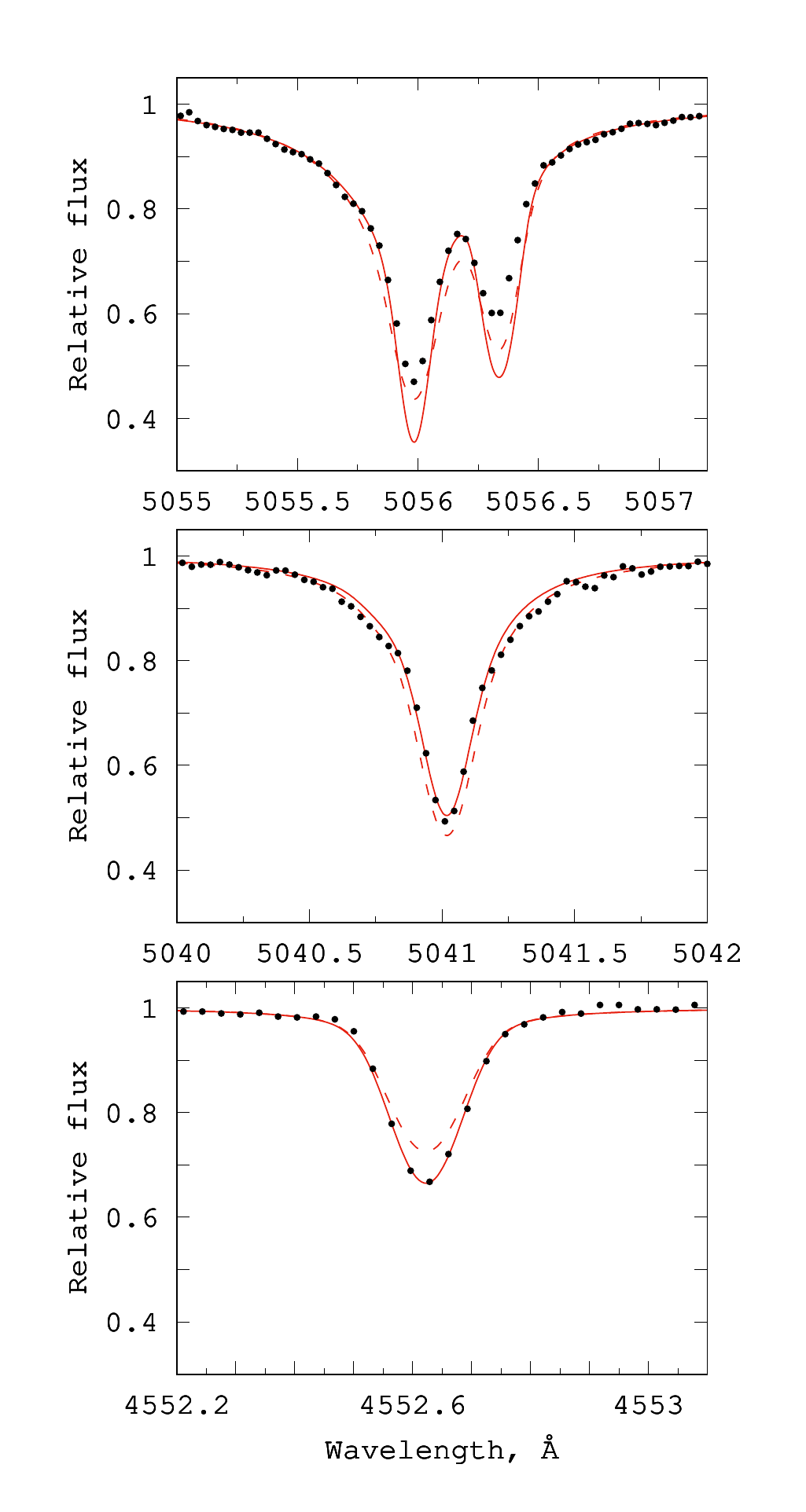}

\caption{\ion{Si}{II} 5055-5056~\AA, 5041~\AA, and \ion{Si}{III} 4552~\AA\ lines in BD+30$\degr$549 (bold dots) compared with the non-LTE (solid curve) and LTE (dashed curve) theoretical profiles. Top panel: the best fits of the \ion{Si}{II} 5056~\AA\ blend obtained with log$N_{Si}/N_{H} = -3.1$ (non-LTE) and $-3.2$ (LTE). Middle and bottom panels: the best non-LTE fits of \ion{Si}{II}5041~\AA\ and \ion{Si}{III} 4552~\AA, respectively. The obtained Si abundances are presented in Table~\ref{tab4}. The same abundances were used to compute the LTE line profiles. }
\label{fig:7}
\end{figure}

The coupled radiative transfer and statistical equilibrium equations were solved with the code {\sc detail} \citep{Giddings81,Butler84}, where the opacity package was updated as presented by
\citet{mash_fe}. Figure~\ref{fig:6} displays the departure coefficients, b$_i = n_i^{\rm NLTE}/n_i^{\rm LTE}$, of the selected levels of \ion{Si}{II} and \ion{Si}{III} involved in the transitions, where the \ion{Si}{II} 5041, 5055, 5056~\AA\ (\eu{4p}{2}{P}{\circ}{} -- \eu{4d}{2}{D}{}{}) and \ion{Si}{III} 4552~\AA\ (\eu{4s}{3}{S}{}{} -- \eu{4p}{3}{P}{\circ}{}) lines arise.
 Here, $n_i^{\rm NLTE}$ and $n_i^{\rm LTE}$ are the SE and thermal (Saha-Boltzmann) number densities, respectively. \ion{Si}{II} is subject to overionization in the line-formation layers, above log~$\tau_{5000}$ = 0, resulting in depleted populations of the ground state and the excited levels up to \eu{4p}{2}{P}{\circ}{} ($E_i$ = 10.1~eV). The upper level of the \ion{Si}{II} \eu{4p}{2}{P}{\circ}{} -- \eu{4d}{2}{D}{}{} transition is depopulated to a lesser extent than is the lower level in the atmospheric layers up to log~$\tau_{5000} \simeq -3$ and, in contrary, to a greater extent in the higher layers. Such a behavior of \eu{4d}{2}{D}{}{} is explained by a competition of the pumping UV transition from the ground state and spontaneous transitions to the levels below \eu{4d}{2}{D}{}{}. As a result, the \ion{Si}{II} 5041~\AA\ line, which forms downwards log~$\tau_{5000} \simeq -3$, is weakened compared with its LTE strength (Fig.~\ref{fig:7}, middle panel) due to b(\eu{4p}{2}{P}{\circ}{}) $< 1$ and the line source function, $S_\nu$, greater than the Planck function, $B_\nu (T)$. The core of \ion{Si}{II} 5055.98~\AA\ forms around log~$\tau_{5000} \simeq -3.4$, where $S_\nu < B_\nu (T)$, and this prevails over b(\eu{4p}{2}{P}{\circ}{}) $< 1$, resulting in strengthened line core (Fig.~\ref{fig:7}, top panel). The wings of \ion{Si}{II} 5055.98~\AA\ form in deeper layers and are weakened compared with the LTE case. Note that, in Fig.~\ref{fig:7}, the non-LTE profile of \ion{Si}{II} 5055.98~\AA\ was computed with a higher Si abundance than that for the LTE profile.
 
 The \ion{Si}{III} levels have enhanced populations in the line-formation layers, resulting in strengthened \ion{Si}{III} lines compared with their LTE strengths, as shown in Fig.~\ref{fig:7} for \ion{Si}{III} 4552~\AA.
 
%%%%%%%%%%%%%%%%%%%%%%%%%%%%%%%%%%%%%%%%%%%%%%%%%%%%%%%%%%%%%%%
%%%%%%%%%%%%%%%%%%%%%%%%%%%%%%%%%%%%%%%%%%%%%%%%%%%%%%%%%%%%%%
\begin{table*}
	\caption{LTE and non-LTE abundances of BD+30$\degr$549 from the \ion{Si}{II}-\ion{}{III} and \ion{Ca}{II} lines}
	\label{tab4}
	\begin{tabular}{lccccccc}
		\hline
		Ion & $\lambda$, \AA & Comment & $E_i$, eV & $\log (gf)$ & \multicolumn{2}{c}{$\log (N_{el}/N_{H})$} &  $\log \tau_{5000}$ \\
		\cline{6-7}
		    &                &         &  &             &  LTE & non-LTE & (line center) \\
		\hline
		\hline
		\ion{Si}{II} & 4621.42 &       & 12.53 & -0.610 & -3.09 & -2.97 & -1.69 \\
		\ion{Si}{II} & 4621.70 & blend & 12.53 & -1.750 & -3.09 & -3.07 & -1.83 \\
		\ion{Si}{II} & 4621.72 & blend & 12.53 & -0.450 & -3.09 & -3.07 & -1.83 \\
		\ion{Si}{II} & 5041.02 &       & 10.07 &  0.030 & -3.15 & -2.97 & -3.23 \\
		\ion{Si}{II} & 5055.98 & $r \ge 0.6$ & 10.07 &  0.520 & -3.20 & -3.10 & -3.56 \\
		\ion{Si}{II} & 5466.85 & blend & 12.52 & -1.380 & -3.20 & -2.99 & -2.26 \\
		\ion{Si}{II} & 5466.89 & blend & 12.52 & -0.080 & -3.20 & -2.99 & -2.26 \\
		\ion{Si}{II} & 6371.37 & $r \ge 0.9$ &  8.12 & -0.080 & -3.18 & -3.18 & -4.07 \\
		mean         &         &       &       &       & -3.15$\pm$0.05 & -3.04$\pm$0.08 & \\		
		\ion{Si}{III} & 4552.62 &      & 19.02 & 0.290 & -2.28 & -2.52 & -0.78 \\
		\ion{Si}{III} & 4567.84 &      & 19.02 & -0.070 & -2.20 & -2.38 & -0.66 \\
		\ion{Si}{III} & 4574.76 &      & 19.02 & -0.410 & -2.51 & -2.62 & -0.56 \\
		mean         &         &       &       &       & -2.33$\pm$0.16 & -2.51$\pm$0.12 & \\
		\ion{Ca}{II} & 5001.48 &      & 7.51 & -0.507 & -4.94 & -4.48 & -0.76 \\
		\ion{Ca}{II} & 5019.97 &      & 7.51 & -0.247 & -5.20 & -4.73 & -0.76 \\
		\ion{Ca}{II} & 5307.22 &      & 7.51 & -0.848 & -4.99 & -4.53 & -0.37 \\
		\ion{Ca}{II} & 5339.19 &      & 8.44 & -0.079 & -5.01 & -4.55 & -0.51 \\
		mean         &         &      &      &       & -5.04$\pm$0.11 & -4.57$\pm$0.11 & \\				
		\hline
	\end{tabular}
\end{table*}    

The LTE and non-LTE synthetic spectra were calculated with the {\sc Synth}V\_NLTE code \citep{Tsymbal_2019}, which implements the pre-computed departure coefficients from the {\sc DETAIL} code. The best fit to the observed spectrum was obtained automatically using the {\sc IDL BinMag6} tool. In the fitting procedure, $\xi_t$ = 0, $V\sin i$ = 0, and $R$ = 48\,000 were fixed, while the Si abundance and macroturbulent velocity were allowed to vary.
 
We found that profiles of the \ion{Si}{II} 4621.4, 4621.7, 5041.0, 5466.8~\AA\ lines are well fitted in non-LTE (see Fig.~\ref{fig:7} for \ion{Si}{II} 5041~\AA) and the obtained non-LTE abundances are higher than the LTE ones, by 0.12 to 0.21~dex (Tab.~\ref{tab4}). For the \ion{Si}{II} 5056~\AA\ blend, the non-LTE profile reproduces the observed one better than the LTE profile, however, fails to fit to the observed line core (Fig.~\ref{fig:7}). The non-LTE abundance obtained from the line wings beyond the relative flux of $r \simeq 0.6$ is only about 0.1~dex lower compared with  that for the well-fitted \ion{Si}{II} lines. The strongest \ion{Si}{II} 6347.1 and 6371.3~\AA\ lines, with their cores formed in the uppermost atmospheric layers (log~$\tau \simeq -4.2$), can only be fitted in their outer wings ($r \gtrsim 0.9$).

The \ion{Si}{III} line profiles in BD+30$\degr$549 are well fitted in the non-LTE calculations. Their non-LTE abundance corrections, $\Delta_{\rm NLTE} = \log A_{\rm non-LTE} - \log A_{\rm LTE}$, are negative and amount to $-0.24$~dex to $-0.11$~dex  (Tab.~\ref{tab4}). 
Thus, non-LTE reduces an abundance discrepancy between \ion{Si}{III} and \ion{Si}{II}, however, it is still substantial, $\log A$(\ion{Si}{III} -- \ion{Si}{II}) = 0.53~dex. The non-LTE calculations confirm a strong enhancement of silicon in the atmosphere of BD+30$\degr$549, with [Si/H] = 1.45 and 1.98 from lines of \ion{Si}{II} and \ion{Si}{III}, respectively.

Our spectrum of BD+30$\degr$549 covers only weak \ion{Ca}{II} lines listed in Table~\ref{tab4}. They all form in deep atmospheric layers and, in the LTE analysis, indicate an enhanced Ca abundance. We performed the non-LTE calculations using the model atom treated by \citet{mash_ca}. It was successfully applied by \citet{Sitnova_2018} to achieve the \ion{Ca}{I}/\ion{Ca}{II} ionization equilibrium in the sample of A-B type stars, including the star $\pi$~Cet with $T_{\rm eff}$ and log~$g$ close to the corresponding parameters of BD+30$\degr$549.

We obtained that, in the line-formation layers of BD+30$\degr$549, \ion{Ca}{II} is subject to overionization, resulting in depleted level populations, weakened spectral lines, and positive non-LTE abundance corrections. The lines under investigation have very similar $\Delta_{\rm NLTE}$ = 0.46 and 0.47~dex (Tab.~\ref{tab4}). Thus, the derived Ca abundance is pushed up to [Ca/H] = 1.11.

The non-LTE calculations were also performed for \ion{Mg}{I}-\ion{Mg}{II} using the non-LTE method treated by \citet{Mashonkina_2018}. In contrast to Si and Ca, Mg is underabundant in the atmosphere of BD+30$\degr$549, with [Mg/H] $\sim -1$ from the LTE analysis. Therefore, compared with the \ion{Si}{II} lines of similar excitation energy and $gf$-value, the \ion{Mg}{II} lines form deeper in the atmosphere, and the non-LTE effects are expected to be smaller. The core of the strongest of the used lines, \ion{Mg}{II} 4481~\AA, forms around log~$\tau_{5000} = -1.8$ and the remaining \ion{Mg}{II} lines listed in Table~\ref{tab3} form close to log~$\tau_{5000} = 0$. Similarly to \ion{Si}{II} and \ion{Ca}{II}, \ion{Mg}{II} is subject to overionization in the atmosphere of BD+30$\degr$549. The non-LTE effects are minor, with $\Delta_{\rm NLTE} <$ 0.01~dex, for all the \ion{Mg}{II} lines except \ion{Mg}{II} 4481~\AA. For the latter, non-LTE leads to the strengthened line core, but the weakened line wings and $\Delta_{\rm NLTE} = -0.06$~dex.

In summary, accounting for the non-LTE effects leads to much better representation of those silicon lines, which are formed at intermediate optical depths (log~$\tau_{5000} > -3$), but fails to reproduce profiles of the strongest \ion{Si}{II} 5056, 6347, 6371~\AA\ lines. Non-LTE reduces, but does not remove the abundance discrepancy between \ion{Si}{II} and \ion{Si}{III}. The obtained results lead us to suspect a presence of a vertical abundance gradient for silicon, with increasing abundance towards deeper layers. We consider the chemical stratification in the next section.

\subsection{Vertical abundance stratification in BD+30$\degr$549 atmosphere}\label{sect:strat}

Abundance analysis of BD+30$\degr$549 under assumption of a chemically homogeneous atmosphere led to the following discrepancies, which are canonical signatures of a vertical abundance gradients \citep{Ryabchikova_2003}:
\begin{itemize}
 \item Strong \ion{Si}{} and \ion{Mg}{} lines require different abundances to fit their wings and cores.\\
 \item The abundances obtained from the lines of the same ion, e.g. \ion{Fe}{II} show a dependence on the excitation energies. The strong \ion{Fe}{II} lines which are formed in the upper layers indicate a lower abundance than the weaker ones, forming deeper in the atmosphere.\\
 \item The lines of the second ions \ion{Fe}{III} and \ion{Si}{III} are abnormally strong and could not be fitted with the same abundances as for the first ions of iron and silicon.
\end{itemize}

Therefore we performed the stratification analysis using the approximation of the vertical abundance distribution in the atmosphere by the step-like function \citep[see e.g][for validation of the method]{Ryabchikova_2005,Kochukhov_2006b}. Stratification calculations were made for three elements \ion{Fe}{}, \ion{Si}{} and \ion{Mg}{} with DDAFit code \citep{Kochukhov_2006b}. The automatic procedure is based on the least-square fitting of the observed line profiles varying the four parameters which characterize the stratification profile: lower and upper abundances, position and width of the abundance jump in the $\log \tau_{5000}$ scale. Selection of spectral lines and atomic data for analysis is of critical importance for reliable reconstruction of stratification profile. The employed linelist should include
lines with different excitation energy of the lower level, $E_i$, and formed at different optical depths, which are uniformly distributed through the atmospheric layers contributing to the line absorption.

%%%%%%%%%%%%%%%%%%%%%%%%%%%%%%%%%%%%%%%%%%%%%%%%%%%%%%%%%%%%%%%%%%%%%%%%%%%%%%%
\begin{figure}
	\includegraphics[width=1.0\linewidth]{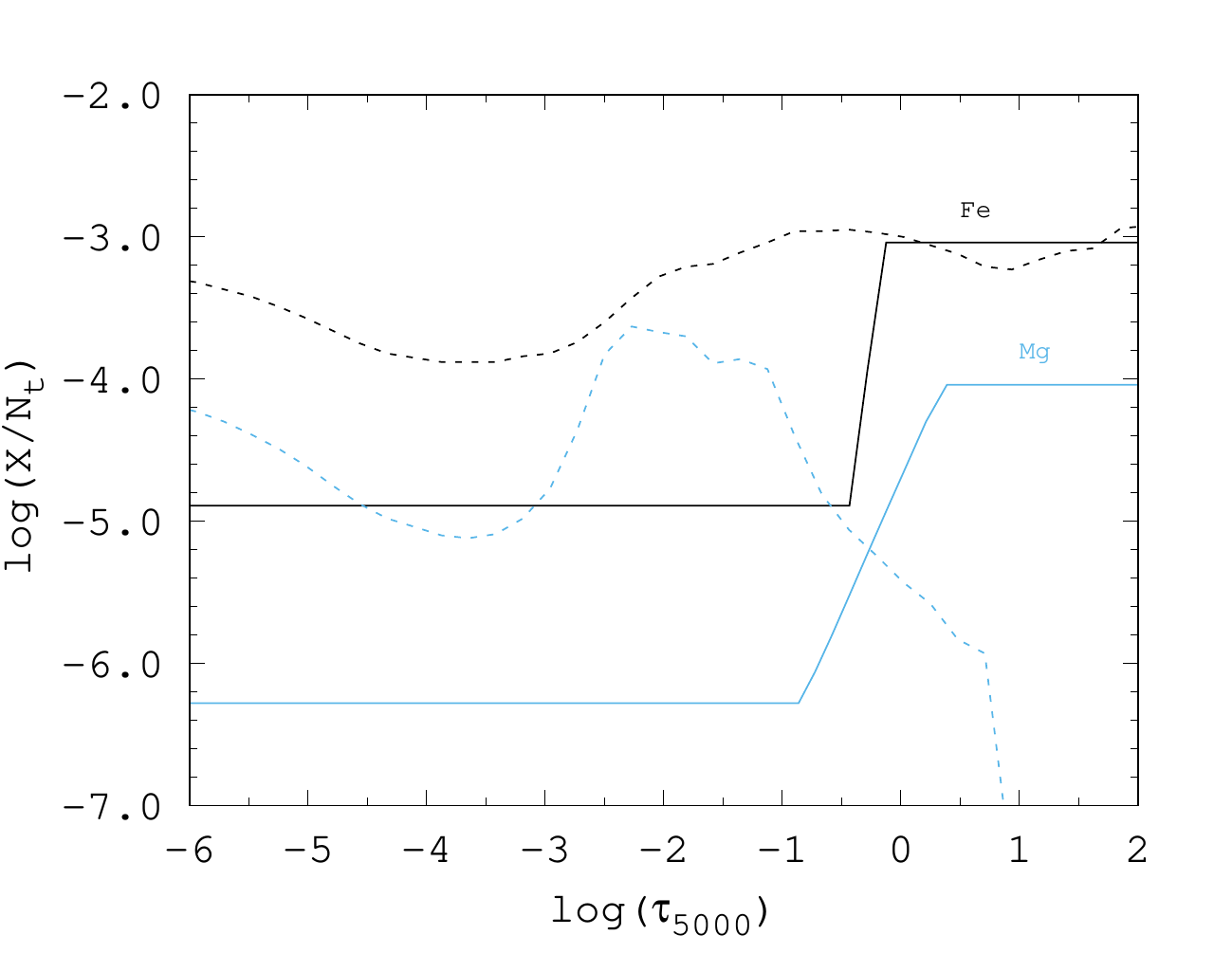}
    \caption{\ion{Fe}{} and \ion{Mg}{} stratification profiles in the BD+30$\degr$549 atmosphere. Results of theoretical diffusion calculations from \citet{LeBlanc_2009} for the same elements at \te=12000~K are shown for reference (dotted curves).}
    \label{fig:8}
\end{figure}
%%%%%%%%%%%%%%%%%%%%%%%%%%%%%%%%%%%%%%%%%%%%%%%%%%%%%%%%%%%%%%%%%%%%%%%%%%%%%%%

The linelist with atomic parameters used in our calculations is presented in Table \ref{tab3}. Thanks to the large number of observed lines, including those from the second ionization stage, these line selection criteria are easily fulfilled for the iron. For iron we performed calculations with two sets of atomic data. First was based on Kurucz's 2010 and 2013 lists for \ion{Fe}{III} and \ion{Fe}{II} respectively, while in second one we used primarily \citet{Raassen_1998} data for \ion{Fe}{II}. Both datasets resulted in the fairly close parameters of the stratification profile, but linelist based entirely on the Kurucz's data made it possible to fit the observed line profiles more accurately. The silicon lines we used also originate from the two ionization states and are more or less evenly distributed in the 8-19 eV energy range. In contrast, most of the available magnesium transitions are in the narrow range of excitation energies. However, the strong \ion{Mg}{II} 4481\AA\, line is formed over a wide range of optical depths  $-1.7 \lesssim \log \tau_{5000} \lesssim 0.8$ and allows to account for the contribution of different atmospheric layers.

\begin{table}
\caption{Linelist used for stratification analysis}
\label{tab3}
\begin{tabular}{lccccc}
\hline
Ion & Wavelength, \AA\, & $E_i$, eV & $\log gf$ & $\log \Gamma_4$ & Ref. \\
\hline
\hline
\ion{Fe}{II} & 4508.280 & 2.85 & -2.420 & -6.530 & K13, WS($gf$) \\
\ion{Fe}{II} & 5018.436 & 2.89 & -1.399 & -6.583 & K13 \\
\ion{Fe}{II} & 5022.418 & 10.35 & -0.054 & -5.567 & K13 \\
\ion{Fe}{II} & 5022.789 & 10.35 & -0.005 & -5.367 & K13 \\
\ion{Fe}{II} & 5030.632 & 10.29 & 0.381 & -5.891 & K13 \\
\ion{Fe}{II} & 5045.106 & 10.31 & -0.151 & -4.984 & K13 \\
\ion{Fe}{II} & 5127.961 & 5.57 & -2.397 & -6.520 & K13 \\
\ion{Fe}{II} & 5169.028 & 2.89 & -1.300 & -6.590 & K13 \\
\ion{Fe}{II} & 5278.939 & 5.91 & -2.520 & -6.696 & K13 \\
\ion{Fe}{II} & 5291.661 & 10.48 & 0.561 & -5.468 & K13 \\
\ion{Fe}{II} & 5303.393 & 8.18 & -1.625 & -5.822 & K13 \\
\ion{Fe}{II} & 5325.552 & 3.22 & -3.185 & -6.603 & K13 \\
\ion{Fe}{II} & 5549.000 & 10.52 & -0.186 & -5.330 & K13 \\
\ion{Fe}{II} & 5567.831 & 6.73 & -1.866 & -6.578 & K13 \\
\ion{Fe}{II} & 6149.246 & 3.89 & -2.732 & -6.588 & K13 \\
\ion{Fe}{III} & 4395.751 & 8.26 & -2.586 & -6.680 & K10 \\
\ion{Fe}{III} & 4419.596 & 8.24 & -1.690 & -6.680 & K10 \\
\ion{Fe}{III} & 5063.467 & 8.65 & -2.922 & -6.680 & K10 \\
\ion{Fe}{III} & 5086.706 & 8.66 & -2.563 & -6.680 & K10 \\
\ion{Fe}{III} & 5114.606 & 8.65 & -3.235 & -6.680 & K10 \\
\ion{Fe}{III} & 5156.111 & 8.64 & -1.922 & -6.680 & K10 \\
\ion{Fe}{III} & 5276.476 & 18.26 & -0.067 & -6.350 & K10 \\
\ion{Fe}{III} & 5282.297 & 18.26 & 0.044 & -6.350 & K10 \\
\\
\ion{Mg}{II} & 4384.637 & 9.99 & -0.790 & -4.07 & NIST \\
\ion{Mg}{II} & 4390.514 & 9.99 & -1.490 & -4.07 & NIST \\
\ion{Mg}{II} & 4390.572 & 9.99 & -0.530 & -4.07 & NIST \\
\ion{Mg}{II} & 4427.994 & 9.99 & -1.208 & -4.40 & NIST \\
\ion{Mg}{II} & 4433.988 & 9.99 & -0.910 & -4.40 & KP \\
\ion{Mg}{II} & 4481.126 & 8.86 & 0.749 & -4.70 & NIST \\
\ion{Mg}{II} & 4481.150 & 8.86 & -0.560 & -4.70 & NIST \\
\ion{Mg}{II} & 4481.325 & 8.86 & 0.590 & -4.70 & NIST \\
\ion{Mg}{II} & 6545.942 & 11.63 & 0.040 & -2.98 & KP \\
\ion{Mg}{II} & 6545.994 & 11.63 & 0.150 & -2.98 & KP \\
\\
\ion{Si}{II} & 4621.418 & 12.52 & -0.610 & -3.86 & M95 \\
\ion{Si}{II} & 4621.696 & 12.52 & -1.750 & -3.86 & M95 \\
\ion{Si}{II} & 4621.722 & 12.52 & -0.450 & -3.86 & M95 \\
\ion{Si}{II} & 5041.024 & 10.07 & 0.030 & -4.80 & M01 \\
\ion{Si}{II} & 5055.984 & 10.07 & 0.520 & -4.76 & M01 \\
\ion{Si}{II} & 5056.317 & 10.07 & -0.490 & -4.76 & M01 \\
\ion{Si}{II} & 5056.317 & 10.07 & -0.490 & -4.76 & M01 \\
\ion{Si}{II} & 5462.144 & 12.88 & -1.107 & -4.04 & K14 \\
\ion{Si}{II} & 5466.460 & 12.52 & -0.080 & -4.20 & M95 \\
\ion{Si}{II} & 5466.849 & 12.52 & -1.380 & -4.20 & M95 \\
\ion{Si}{II} & 5466.894 & 12.52 & -0.080 & -4.20 & M95 \\
\ion{Si}{II} & 5469.451 & 12.88 & -0.762 & -4.06 & M95 \\
\ion{Si}{II} & 5469.233 & 16.73 & -1.100 & -4.06 & M95 \\
\ion{Si}{II} & 6371.371 & 8.12 & -0.080 & -5.08 & M01 \\
\ion{Si}{III} & 4552.622 & 19.02 & 0.290 & 0.00 & NIST \\
\ion{Si}{III} & 4567.840 & 19.02 & -0.070 & 0.00 & NIST \\
\ion{Si}{III} & 4574.757 & 19.02 & -0.410 & 0.00 & NIST \\
\hline
\end{tabular}
\bigskip

\emph{References:} K10,K13,K14 - Kurucz' online database of observed and predicted atomic transitions (http://kurucz.harvard.edu/atoms/); WS - \citet{Den_Hartog_2014} ; KP - \citet{KP_1975}; NIST - \citet{NIST_ASD}; M95 - \citet{Mendoza_1995}; M01 - \citet{Matheron_2001}. \\
\end{table}    

\begin{figure*}
	\includegraphics[width=1.0\linewidth]{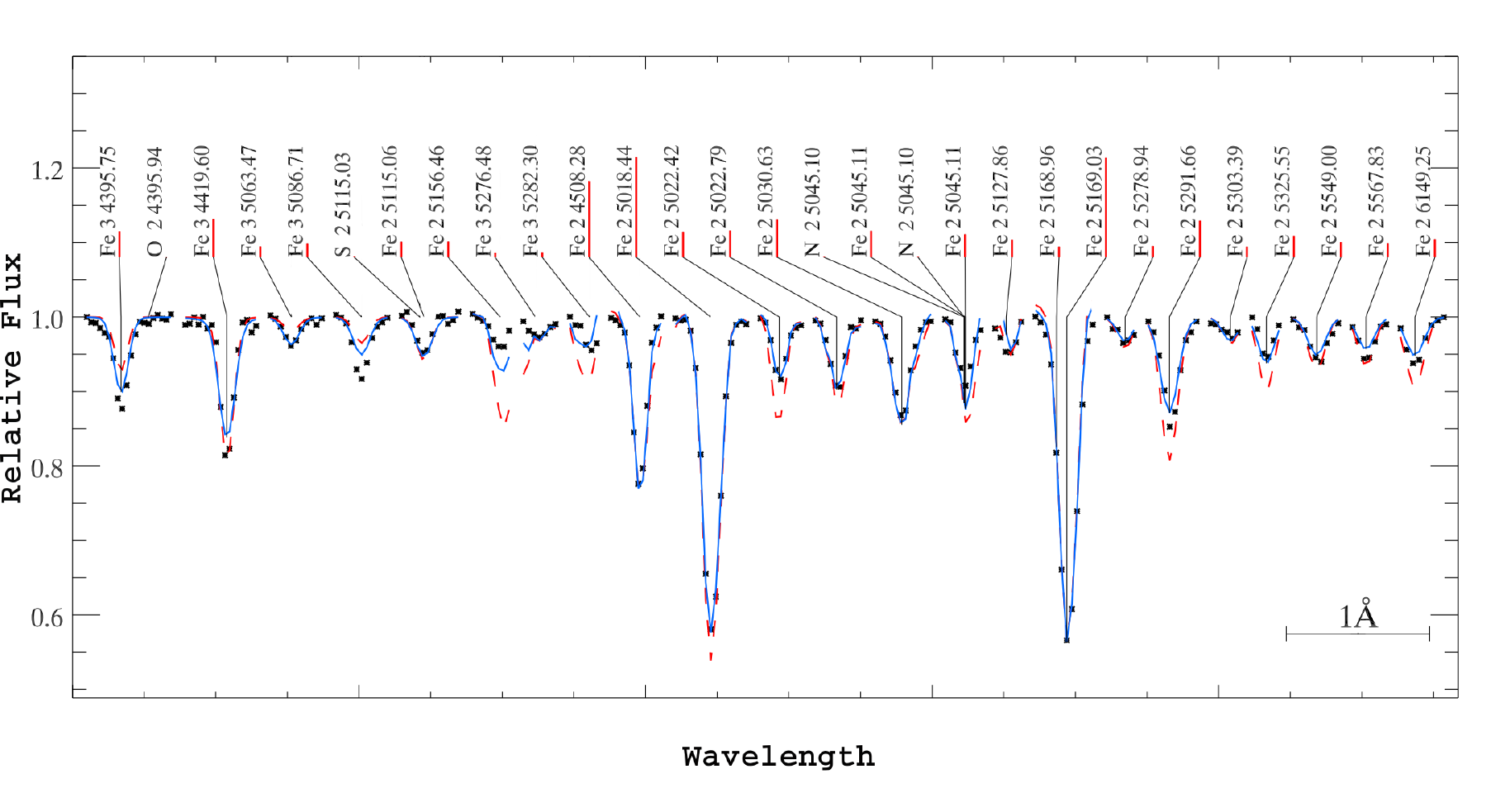}
    \caption{Observed \ion{Fe}{} line profiles (shown by dots) in comparison with synthetic spectra calculated with homogeneous abundance distribution (dashed curves) and with vertical stratification (solid curve).}
    \label{fig:9}
\end{figure*}

\begin{figure}
	\includegraphics[width=0.7\linewidth]{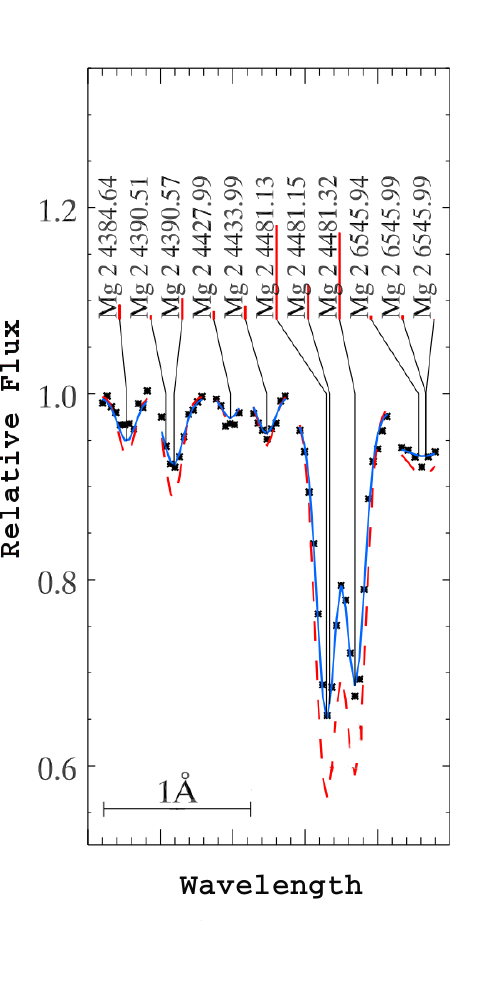}
    \caption{The same as Fig. \ref{fig:9} for \ion{Mg}{} lines}
    \label{fig:10}
\end{figure}

The resulting abundance stratification for \ion{Mg}{} and \ion{Fe}{}  are shown in Fig. \ref{fig:8}. The iron and magnesium possess the step abundance gradient in the narrow range of optical depths near $\log \tau_{5000}\approx -0.3$. Both elements show tendency to increase abundance in the deeper atmosphere. Iron is mildly depleted in the upper atmosphere and increased its concentration with depth up to $\sim1.5$ dex excess relative to the Sun. Magnesium is significantly underabundant by 1.9 dex in the upper atmospheric layers, reaches the solar value at $\log \tau_{5000}\approx0$ and experiences a mild overabundance in the deepest layers contributing to line absorption. Comparison between observed lines profiles and synthetic ones calculated with stratified abundances shows a reasonable agreement in case of \ion{Fe}{} and \ion{Mg}{}. Compared with the homogeneous distribution, accounting for stratification allows to adequately reproduce
the line core depths for \ion{Fe}{} lines arising from both ionization stages (Fig. \ref{fig:9}), as well as the profiles of magnesium lines including wings of the \ion{Mg}{II} 4481\AA\, line (Fig. \ref{fig:10}). 

In contrast, the jump of stratification profile for silicon turned out to be very gradual, distributed over the wide range of optical depths. Resulting fit of the observed lines profiles is poor and can reproduce neither the enhanced absorption in the \ion{Si}{III} lines, nor the wings of the strong  \ion{Si}{II} lines.

%%%%%%%%%%%%%%%%%%%%%%%%%%%%%%%%%%%%%%%%%%%%%%%%%%%%%%%%%%%%%%%%%%%%%%%%%%%%%%%
%%%%%%%%%%%%%%%%%%%%%%%%%%%%%%%%%%%%%%%%%%%%%%%%%%%%%%%%%%%%%%%%%%%%%%%%%%%%%%%

%%%%%%%%%%%%%%%%%%%%%%%%%%%%%%%%%%%%%%%%%%%%%%%%%%%%%%%%%%%%%%%%%%%%%%%%%%%%%%%
%%%%%%%%%%%%%%%%%%%%%%%%%%%%%%%%%%%%%%%%%%%%%%%%%%%%%%%%%%%%%%%%%%%%%%%%%%%%%%%

Qualitatively, depth-dependence of iron and magnesium abundance as well as the depletion of helium in the upper atmosphere of BD+30$\degr$549, is consistent with the results of theoretical calculations \citep[][]{LeBlanc_2009} and points to selective diffusion as the reason for the vertical stratification of these elements. However, exact position and magnitude of the abundance jumps in the BD+30$\degr$549 atmosphere deviates from the model predictions. First, these parameters depend on the physical conditions in a given atmosphere, and, second, the calculations represent the equilibrium solution while equilibrium could not yet be reached in the young BD+30$\degr$549 atmosphere.
It is interesting to note that the results of the \citet{LeBlanc_2009} calculations show a flattening of the stratification profiles as the \te\ rises from 8000~K to 12000~K. The latter value is the maximum effective temperature used in their calculations, and the resulting abundance gradients are overplotted in our Fig. \ref{fig:8}. It is evident that the stratification profiles in the hotter atmosphere of BD+30$\degr$549 are steeper, in contrast to the theoretical predictions. This can be due to the star's young age and non-equilibrium diffusion processes.

\subsection{Impact of the \ion{Si}{} overabundance on the atmospheric structure}\label{sect:si_impact}

Our abundance analysis revealed that both silicon and iron are substantially overabundant in BD+30$\degr$549 atmosphere (Sect. \ref{sect:atm_param}) and also these elements possess both vertical and probably lateral abundance gradients (Sect. \ref{sect:strat}; \ref{sect:phot_var}). These metals play the important role in the opacity distribution throughout the atmosphere due to bound-free transitions and consequently affect its thermal balance \citep[e.g.][]{Khan_2007}. Indeed, as far back as \citet{Strom_1969} it was shown that $\approx$1.5~dex silicon enhancement affects the structure of the atmosphere in the same way as a $\sim$1000~K increase in \te~ and also causes a flux redistribution in the UV.

\begin{figure}
	\includegraphics[width=0.99\linewidth]{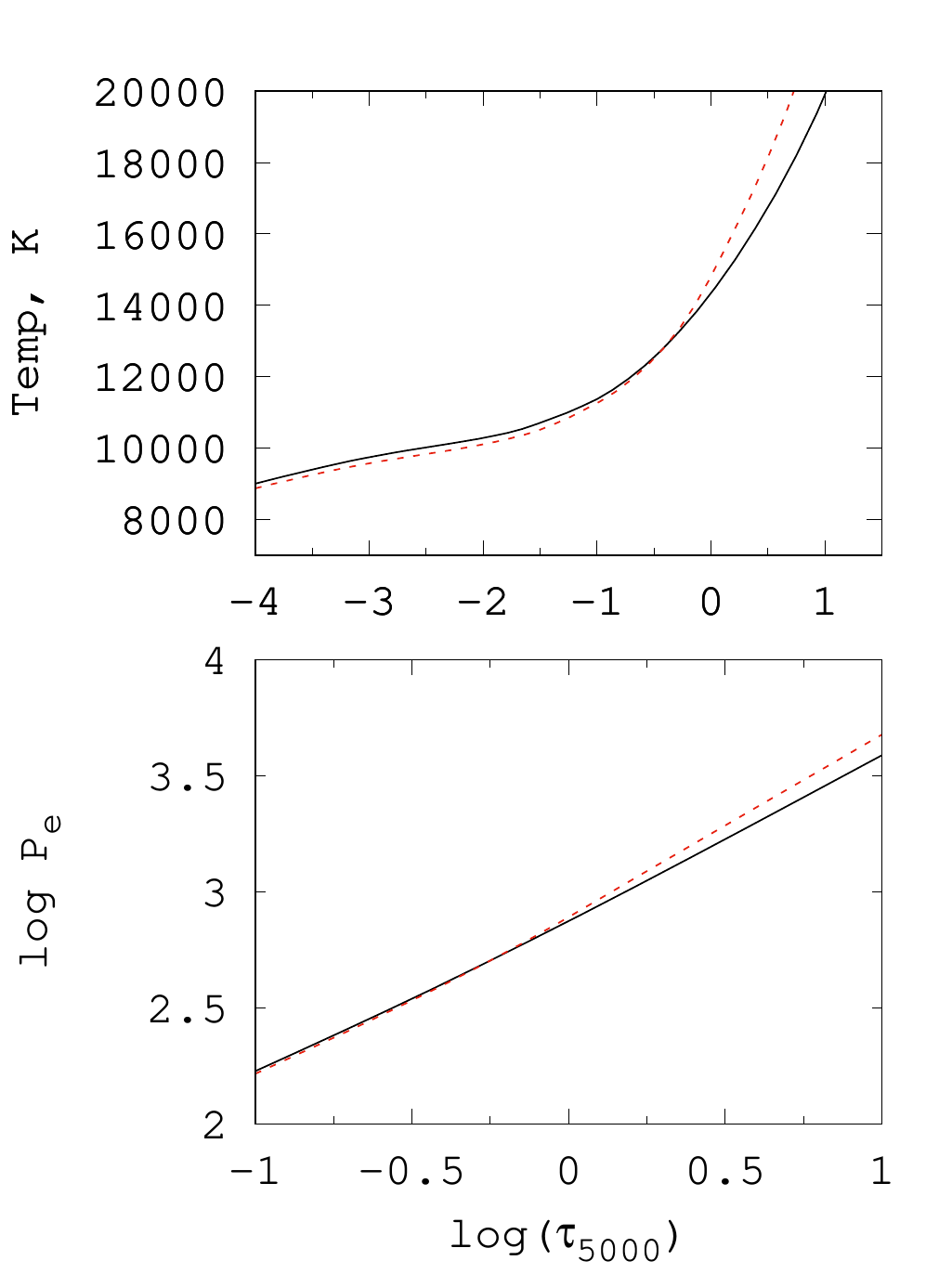}
    \caption{Dependence of temperature (top panel) and electronic pressure (bottom panel) on the optical depth for the helium-weak and silicon-rich homogeneous (shown by solid curve) and stratified (dashed curve) models.}
    \label{fig:11}
\end{figure}

We performed a preliminary series of calculations to check how the vertical and horizontal stratification of the silicon affects the atmospheric model structure and emergent spectrum in case of BD+30$\degr$549. First, we calculated with \textsc{LLmodel} code the atmospheric model taking into account the individual abundances determined in BD+30$\degr$549 atmosphere as well as the vertical stratification for \ion{Si}{}, \ion{Fe}{} and \ion{Mg}{}. For iron and magnesium we applied stratification profiles derived from our analysis (see Fig. \ref{fig:8}). For silicon profile was set manually. The abundance $\log (N_{\ion{Si}{}}/N_{H})=-2.33$~dex deduced from the LTE determination using \ion{Si}{III} lines was adopted for the deep atmospheric layers, while as the upper limit we took $\log (N_{\ion{Si}{}}/N_{H})=-3.76$~dex obtained from our fitting procedure with \textsc{DDAFit}. The location of the abundance jump was set at the same optical depths as for iron. Results of calculations are shown in Fig. \ref{fig:11}. One can see that the temperature structure of the model changed noticeably both in the regions of lines and continuum formation. In the line forming region, we obtained a trend of decreasing temperature with depth compared to the homogeneous model (also helium-weak and silicon-rich). The difference reaches $\Delta T\approx$3$\%$ at $\log \tau_{5000}=-1$. The change in the electron pressure was less pronounced and occured predominantly in the deep atmospheric layers with $\log \tau_{5000}\gtrsim0$. The joint effect of these changes on the SED and line profiles appeared to be significant. The stratified model provides the larger Balmer discontinuity and steeper Balmer continuum in UV. The latter resulted in better fit of the observed fluxes in the XMM-OM $UVW2$ and $UVM2$ filters than in case of homogeneous model (see Fig. \ref{fig:2}). The magnitude of the Balmer jump produced with stratified model is less consistent with the observations, but this can be compensated by small variations in extinction and increasing \te~ within $\sim$300~K. Nevertheless, in the synthetic spectrum calculated with the stratified model, the Balmer lines became much broader. Fitting the theoretical profiles to the observed ones requires significant temperature correction, conflicting with the SED fitting. We conclude that current calculations with the homogeneous provide better simultaneous representation of the SED and line absorption spectrum of BD+30$\degr$549.

%%%%%%%%%%%%%%%%%%%%%%%%%%%%%%%%%%%%%%%%%%%%%%%%%%%%%%%%%%%%%%%%%%%%%%%%%%%%%%%
%%%%%%%%%%%%%%%%%%%%%%%%%%%%%%%%%%%%%%%%%%%%%%%%%%%%%%%%%%%%%%%%%%%%%%%%%%%%%%%
  
Although our spectrum allowed to put only an upper limit on the strength of the magnetic field of BD+30$\degr$549 at the date of observation, generally helium-weak silicon stars belongs to the magnetic sequence of CP stars \citep{Romanyuk_2007}. Therefore, if BD+30$\degr$549 hosts a large-scale magnetic field, it could also lead to the patchy horizontal distribution of elements, i.e. existence of elemental spots with an altered temperature structure. The detected photometric variability of BD+30$\degr$549 preliminary confirms this assumption. 

To take into account the influence of the silicon spot on the emergent spectrum, we used approach similar to that applied for another silicon star CU Vir \citep{Krivoseina_1980}. A grid of simple models in which $\sim$50\% of the stellar surface is occupied by a spot with enhanced silicon abundance and the rest of the surface being with the solar abundance has been computed. The temperature structure of the spot was recomputed with the \textsc{LLmodel} code, taking into account the opacity due to the increased silicon abundance. Varying the spot area and silicon abundance within reasonable limits, we were able to reproduce the profiles of individual silicon lines. However, the attempt to simultaneously reproduce \ion{Si}{II} and \ion{Si}{III} lines with the same abundance in the spot also failed. A simple estimate also shows that presence of such a spot with [\ion{Si}{}/H]=+1.8~dex leads to rotational brightness modulation with an amplitude up to $\sim0.1^m$ magnitude in the $V$ filter. This value is less than the observed $\Delta V\approx0.2^m$ amplitude, but it should be taken into account that in reality the temperature structure of the spot could be modified by opacity due to other elements as well as by effects of vertical stratification. Thus, in order to reproduce the emergent spectrum, one needs to know both the distribution of spots on the stellar surface and the elemental abundances within each of them. Such information can be obtained from a subsequent spectroscopic monitoring of BD+30$\degr$549.

%%%%%%%%%%%%%%%%%%%%%%%%%%%%%%%%%%%%%%%%%%%%%%%%%%%%%%%%%%%%%%%%%%%%%%%%
\section{Discussion}
%%%%%%%%%%%%%%%%%%%%%%%%%%%%%%%%%%%%%%%%%%%%%%%%%%%%%%%%%%%%%%%%%%%%%%%%
\subsection{Stellar parameters, chemical abundances and stratification}

Our spectroscopic analysis of BD+30$\degr$549 atmosphere yields the following parameters: \te=13100$\pm$100~K and \logg = 4.2$\pm$0.1. According to the modern temperature calibrations of MKK system \citep{Gray_Corbally_2009,Pecaut_Mamajek_2013} such an effective temperature, as well as the bolometric absolute magnitude $M_\mathrm{bol}^{*}=-0.5^m$, are in the reasonable agreement with the historical spectral classification of the star as B8. Our analysis revealed the significant depletion of helium in the atmosphere of BD+30$\degr$549 with $\log (A)_{\ion{He}{}}\leq -3.5$ dex. Other elements on average are also depleted, with the exception of \ion{Ca}{}, \ion{P}{}, \ion{Si}{} and \ion{Fe}{} which possess mild- to strong overabundance. The helium-weak stars constitute a rather heterogeneous group of hot CP stars and reveal a diversity of the element abundance patterns \citep[see e.g.][]{Ghazaryan_2019}. With the obtained element abundance pattern, BD+30$\degr$549 is not an outlier in this group. We attribute BD+30$\degr$549 with a caution to the silicon subgroup of helium-weak stars because of strikingly enhanced lines of \ion{Si}{II} and \ion{Si}{III} in its spectrum.

Except the general enrichment of silicon in BD+30$\degr$549 atmosphere which exceeds more than 1.3~dex with respect to the solar value, the lines corresponding to the two ionization stages yield an abundance difference of about 0.8 dex. In fact, such a discrepancy in abundances deduced from the \ion{Si}{II} and \ion{Si}{III} lines is known for B-type stars \citep[see][and references therein]{Bailey_2013}. The latter authors studied this "\ion{Si}{II}/\ion{}{III}-anomaly"\, on the representative sample containing normal B-type stars as well as magnetic Bp and HgMn stars. It was shown that both magnetic and non-magnetic stars exhibit $\log A$(\ion{Si}{III} -- \ion{Si}{II}) difference, but in case of normal B stars it is less pronounced reaching $\sim0.3-0.8$~dex. The non-LTE effects and abundance stratification were proposed as possible reasons for this "\ion{Si}{II}/\ion{}{III}-anomaly"\,. Indeed the non-LTE calculations by \citet{Mashonkina_2020} resulted in agreement between \ion{Si}{III} and \ion{Si}{II} based abundances for normal B7 and B3 stars, i.e. eliminated the difference in LTE abundances of 0.23 and 0.7~dex.

However, in case of BD+30$\degr$549 even after accounting for the non-LTE effects the $\approx$0.5 dex difference still persists between abundances deduced from the \ion{Si}{II} and \ion{Si}{III} lines. Intuitively, the discrepancy in the silicon lines after accounting for the non-LTE effects suggest its depletion in the upper atmosphere with increasing abundance in the deepest layers. Indeed, we found clear observational evidence of vertical abundance stratification in BD+30$\degr$549 manifested in the lines of some other elements, e.g. \ion{Fe}{} and \ion{Mg}{}. However, in our analysis we were unable to construct such a stratification profile to equalize the abundances determined from the silicon lines forming at significantly different optical depths. Although at present state we cannot fully explain the "\ion{Si}{II}/\ion{}{III}-anomaly"\, observed in BD+30$\degr$549 spectrum, our results clearly indicate that combination of the effects of abundance stratification as well as departures from LTE in the silicon's line formation region contributes to its appearance, in agreement with suggestion by \citet{Bailey_2013}.

Another important factor which have to be accounted is inhomogeneous lateral distribution of elements over the stellar surface (chemical spots), which is typical for Ap/Bp stars. In the case of BD+30$\degr$549, the spots manifest in the rotationally-modulated period, detected in photometric variability of the star. Our preliminary calculations have shown that both silicon overabundance and vertical gradient of its abundance also affects significantly the temperature structure of the atmosphere. Thus, we reinforce the conclusion by \citet{Kochukhov_2018} that sophisticated interpretation of the observed spectra of CP stars generally requires taking into account the 3D distribution of abundances and subsequent recalculation of the atmospheric structure.

\subsection{Evolutionary status and conditions for early development of chemical peculiarity}

CP helium stars are abundant among the young stellar population in OB associations and Galactic star forming regions. For a long time numerous helium-weak stars are known in young Orion and Scorpius-Centaurus OB associations \citep[see review in][]{Smith_1996}. The existence of helium-weak stars in Ori OB1c ($\sim2-6$ Myr old, \citet{Bally_2008b}) subgroup \citep{Romanyuk_2013}, as well as recent finding of the helium-weak silicon star in the Orion's Trapezium region \citep{Costero_2021} (although without unambiguous attribution to $\sim1$ Myr old Trapezium cluster) indicate that the onset of this type of peculiarity is either a very rapid process, or it can be triggered long before star's settling on the ZAMS, during the PMS phase. The target of the present study, BD+30$\degr$549, is the member of active NGC 1333 star forming region and still embedded in the reflection nebula. Its isochronal age $t\approx2.7$ Myr is slightly larger than the median cluster age $\sim1.5$ Myr \citep{Luhman_2016} but is well within age dispersion found by \citet{Aspin_2003}. In the HR diagram, BD+30$\degr$549 lies at the end of its PMS track, very close to the ZAMS. 

Results of modern time-dependent diffusion calculations show that the diffusion timescale in the typical Ap/Bp atmosphere is of order $\sim10^1-10^3$ yr \citep{Alecian_2011,Alecian_2019}, and hence development of the surface chemical inhomogeneities can occurs rapidly when the favorable conditions for an effective atomic diffusion have been achieved. Basically, such conditions are set up by slow axial rotation and hence ineffective meridional circulation as well as by suppressing of the microturbulence.

Indeed, our spectroscopic analysis revealed zero turbulence in the atmosphere of BD+30$\degr$549 and undetectable rotational broadening of the absorption lines with the upper limit on projected rotational velocity $V\sin i\lesssim2$ km$\cdot$s$^{-1}$. Is BD+30$\degr$549 a really slow rotator or it is observed pole-on? If the photometric variability of the star is rotationally-modulated, then the $\approx$123$^d$.3 period corresponds to the equatorial rotation velocity $V_{*}\lesssim1$ km$\cdot$s$^{-1}$, adopting the stellar radius $R_*=2.2R_{\odot}$. This velocity is in good agreement with our spectroscopic determination and implies that the star is observed close to equator-on. If we assume that during the PMS phase, BD+30$\degr$549 followed the evolutionary path of a typical Herbig Ae/Be star, then after the main protostellar mass-accretion phase it should gained the angular momentum resulted in typical equatorial rotational velocity of order $V_*\approx$150 km$\cdot$s$^{-1}$ (corresponding rotational period $\approx0.7^d$ for $R_*\approx2R_{\odot}$). Comparing this value with the present day rotational period we conclude that that BD+30$\degr$549 has lost (or not gained) a significant fraction of its angular momentum during the PMS phase. 

The main agent in various macroscopic mechanisms of rotational braking \citep[see][for review]{Bouvier_2013} is believed to be the magnetic field, which is also effectively suppresses the turbulence and various types of circulation in the outer layers of the atmosphere \citep{Mestel_1977}. Another possibility sometimes invoked to explain the loss of the angular momentum by Ap/Bp stars is synchronization in a binary system. Below we will examine both of these scenarios to see if they can explain an increase the rotational period of BD+30$\degr$549 from $\approx0.7^d$ to about $123^d$ during the lifetime of the star, i.e $\approx2.7$ Myr.

\subsubsection{Synchronization in binary system}

The structure of the SED of BD+30$\degr$549 shows the absence of the NIR emission above the photospheric level within $\approx$1.2-8 $\micron$ range, indicative for the inner cavity in the disk. At the same time, the mid-IR excess peaked at 24 $\micron$ is approximated by $\sim200$~K blackbody radiation and traces warm dust on the inner edge of the disk. Assuming $T_d\backsimeq$200~K as the equilibrium temperature of the dust, we can estimate its radial distance from the relation $R_d/R_*\approx(T_*/T_d)^2$. Substituting the stellar parameters $T_*$=13100~K and $R_*=2.2R_\odot$ we obtain the inner cavity radius $R_d\approx$50 au. Such a radial distribution of dust is typical for evolved transitional or debris disks \citep{Williams_2011}. Recent interferometric observations showed that such large cavities in disks around Herbig Ae/Be stars are widespread \citep[e.g.][]{van_der_Plas_2017a,van_der_Plas_2017b} and could be opened by both accretion and photoevaporation processes, as well as by the orbital motion of the secondary companion.

If we assume that the radius of the circular orbit of the putative secondary companion coincides with the inner boundary of the disk, then corresponding Keplerian period at this distance will be $\approx$196 yr for the given stellar parameters. However, the variability of the mid-IR excess indicates the existence of the population of $\sim$1000-km size bodies at this distance rather than single massive companion (see Sect. \ref{sect:phot}). Obviously, such bodies cannot play a role in tidal slowdown of the host star.

Moreover, synchronization for such a long period is fundamentally doubtful. Both the observational studies \citep{Abt_1973,Gerbaldi_1985} and theoretical consideration \citep{Zahn_1977} suggest that synchronization usually occurs for the short orbital periods $P<6^d$, and statistically the number of synchronized systems among Ap/Bp stars is low. Thus, we have to assume the existence of a closer and more massive companion inside the central cavity in the disk in order to explain the braking of BD+30$\degr$549 rotation due to synchronization. Currently neither the photometric behavior of the star, nor the structure of its disk, indicates existence of such a companion. However, we cannot completely rule out its existence and observations with high-angular resolution could be beneficial.

\subsubsection{Magnetic braking}

Another explanation for slowing down the rotation of the BD+30$\degr$549 is magnetic braking. The evolution of the angular momentum $J$ of the disk-bearing PMS star can be generally expressed as:

\begin{equation}
 \frac{dJ}{dt}= T_{acc}+T_{disk}+T_{wind} 
 \label{eqn:1}
\end{equation}
In the right hand part of the equation the terms with specific lower indexes correspond to the torques due to accretion, disk locking and magnetized wind. Generally the accretion term $T_{acc}$ provides the angular momentum inflow and leads to spin up of the star. The remaining terms $T_{disk}$ and $T_{wind}$ describe the braking due to the magnetic coupling with the disk \citep[e.g.][]{Rosen_2012} and carrying away the angular momentum via the magnetized wind \citep{Weber_1967}. For the simplified estimation of the braking timescale of BD+30$\degr$549 during its PMS phase let us consider the action of only braking terms neglecting spin up due to accretion. Before a special comment have to be given on the origin of the wind which we will consider below. The fully radiative early type stars do not host solar-type winds powered by dynamo-generated magnetic field originally employed in Weber\&Davis mechanism. Also the radiatively-driven wind is insufficient within late-B stars temperature domain \citep{Babel_1996} with the mass-loss rates of order $\dot M_{out}<10^{-14}$ $M_{\odot}\cdot$yr$^{-1}$. However during the phase of active accretion there are few types of accretion-driven winds, e.g. "X-wind"\, or "stellar wind"\, which propagate via the open lines of stellar magnetic field \citep[see][for review]{Ferreira_2013}. These winds carrying away the substantial mass up to $\dot M_{out}\approx10^{-8}$ $M_{\odot}\cdot$yr$^{-1}$ and thus could have the sufficient impact on the angular momentum evolution of young star. However, their operational time is limited by the period of active accretion.   

Let us assume that during the significant fraction of its PMS evolution BD+30$\degr$549 followed the typical path of accreting Herbig Ae/Be star. First we consider the disk locking, which should lead to spin down of the stellar rotational rate $\Omega_*$ to the angular velocity $\Omega_K=\sqrt{GM_*/R_d^3}$ at the inner boundary of the Keplerian disk located at distance $R_d$ from the central star. It is usually assumed that the inner boundary of the full accretion disk is determined by the Alfven radius ($R_d=R_A$), i.e radius of magnetosphere. The latter can be calculated using the expression \citep{Lamb_1973}:

\begin{equation}
 \frac{R_A}{R_*}=\frac{B_*^{4/7}R_*^{5/7}}{\dot M_{acc}^{2/7}(2GM_*)^{1/7}}
 \label{eqn:2}
\end{equation}

Where $B_*$ is the strength of dipole component of stellar magnetic field, and $\dot M$ - mass accretion rate. Assuming the typical accretion rate $\dot M\approx10^{-7}$ $M_{\odot}\cdot$yr$^{-1}$, $B_*\approx500$~G field which is characteristic for the magnetic Herbig star \citep{Alecian_2013} and substituting the mass and radius of the star into the Eq.\ref{eqn:2}, we obtain that BD+30$\degr$549 has a very compact magnetosphere $R_A/R_*\approx1.5$. If the field strength matches those of magnetic Ap/Bp stars with $B_*\approx2$~kG, then $R_A/R_*\approx3.3$. Keplerian periods of $\approx0.4^d$ and $1.3^d$ days correspond to these radii. Thus, in the first case, the coupling with the disk should lead to the spin up of the star, while in the second case the deceleration turns out to be insignificant to achieve the present day rotational rate of BD+30$\degr$549. 

On the other hand the timescale $\tau=J/\dot J$ of the angular momentum $J=k^2M_*R^2_*\Omega_*$ loss due to the magnetically-driven mass loss with the rate $\dot M_{out}$ can be found from the equation:       

\begin{equation}
 \frac{dJ}{dt}=\frac{J}{\tau}=\frac{2}{3}\Omega_*\dot M_{out}R_A^2
 \label{eqn:3}
\end{equation}

Using for clarity the largest value $R_A/R_*\approx3.3$, and adopting the relatively high mass-loss rate $\dot M_{out}/\dot M_{acc}\approx0.1$ and parameterized gyration radius $k=1$ we finally obtained from Eq.\ref{eqn:3} the estimation of the braking time of the star from the initial equatorial velocity $V_*=150$ km$\cdot$s$^{-1}$ as  $\tau \approx 40$ Myr. Obviously this value is of order magnitude larger than the duration of PMS evolution of $3.2M_{\odot}$ star and indicate that magnetic wind-braking in its simplest form is not enough efficient to explain the present slow rotation of BD+30$\degr$549. Moreover, the obtained $\tau$ is the low limit since it was implicitly assumed that the mass-loss rate remained constant throughout the whole considered time span. In fact, in addition to the decrease of the accretion rate with time, dissipation of the inner disk in BD+30$\degr$549 (see Sect. \ref{sect:phot}) and the corresponding termination of the accretion/outflow processes indicate that the action of this mechanism also halted before it could lead to a significant deceleration of the star.    

Our simplified estimations are in agreement with those of \citet{Spruit_2018} and with results of more sophisticated numerical calculations by \citet{Rosen_2012} who showed that the spin down of the intermediate mass stars as well as formation of the long-period-tail in rotational period distribution of the Ap/Bp stars is unlikely directed by the mechanisms of disk locking and wind braking considered above. Alternatively, \citet{Spruit_2018} proposed an idea that in rare cases the initial phases of protostellar mass accretion can occur without acquiring of angular momentum. For this scenario of "magnetically dominated accretion"\,, it is important that the connection between the magnetic field of the disk and the field inside the parental cloud is preserved. Application of this model to BD+30$\degr$549 suggests that its long lasting association with reflection nebula could be an essential factor. Indeed the survey of NGC 1333 with James Clerk Maxwell Telescope (JCMT) \citep{Doi_2020} revealed the presence of dust polarized emission in the BD+30$\degr$549 region which is indicative for the particles alignment by the global interstellar (IS) magnetic field. This IS field itself is aligned in NW-SE direction, which is roughly coincidental with orientation of the reflection nebula in the optical images. The IS field may be the anchor that initially prevented the star from gaining large angular momentum at the stage of protostellar collapse. Detection of the proper magnetic field of BD+30$\degr$549 with the spectropolarimetric observations is essential to proof this magnetic braking scenario.

%%%%%%%%%%%%%%%%%%%%%%%%%%%%%%%%%%%%%%%%%%%%%%%%%%%%%%%%%%%%%%%%%%%%%%%%
\section{Conclusions}
%%%%%%%%%%%%%%%%%%%%%%%%%%%%%%%%%%%%%%%%%%%%%%%%%%%%%%%%%%%%%%%%%%%%%%%%
We summarise our main findings as follows:\\
\begin{enumerate}
\item BD+30$\degr$549 is $\approx$2.7 Myr old PMS or early-ZAMS member of NGC 1333 star forming region with pronounced chemical peculiarity of helium-weak silicon type. With the spectral synthesis technique we found the following parameters of its atmosphere \te=13100$\pm$100~K, \logg=4.2$\pm$0.1 and also detected the evidence for the highly stabilized atmosphere with negligible axial rotation and zero turbulence.\\

\item The average chemical composition indicates a deficit of almost all of the investigated elements, except \ion{Si}{}, \ion{Fe}{}, \ion{Ca}{} and \ion{P}{}. The \ion{Si}{} is overabundant up to 2.2 dex, with significant difference in abundance determined with the lines of the first and second ions (so-called "\ion{Si}{II}/\ion{}{III} anomaly"\,). We also found additional observational signatures of vertical abundance stratification in BD+30$\degr$549 atmosphere.\\

\item Non-LTE calculations resulted in minor- to moderate corrections for \ion{Mg}{} and \ion{Ca}{} abundances respectively. Also NLTE approach leads to much better reproduction of individual silicon line profiles, but does not completely remove the abundance discrepancy between \ion{Si}{II} and \ion{}{III}. \\

\item We determined stratification profiles for iron and magnesium, which show depletion of these elements in the upper atmosphere and increased concentrations with depth. We were unable to reconstruct the reliable stratification profile for the silicon. \\

\item  Currently neither the non-LTE effects nor the vertical abundance gradient exclusively explain the "\ion{Si}{II}/\ion{}{III} anomaly"\, observed in BD+30$\degr$549 spectrum. The observed difference in abundances probably could be reduced by accounting also for the non-uniform lateral distribution of the silicon, i.e. its concentration in the spots with an altered temperature structure. The existence of such a spots is suspected on the basis of the $\approx123^d$ period found in low-amplitude photometric variability of BD+30$\degr$549. \\

\item The vertical distribution of elements in BD+30$\degr$549 atmosphere is qualitatively consistent with results of theoretical diffusion calculations and points to atomic diffusion as mechanism of their formation.\\

\item The star with an age $t\approx$2.7 Myr lies close to the ZAMS. Nevertheless, the conditions favorable for development of the peculiar chemical composition under the action of the diffusion processes apparently arose at the PMS phase. One of the mechanisms responsible for the slow rotation and rapid stabilization of the atmosphere could be the sustained magnetic linkage of the stellar field lines with interstellar magnetic field in the parental cloud. \\
\item The detection of the mid-IR excess indicates that BD+30$\degr$549 hosts circumstellar disk with developed inner cavity. Tentatively discovered variability of the flux at 24$\micron$ on the decadal timescale is likely indicative for the ongoing collisions in planetesimal belt producing significant amount of secondary generated dust.

\end{enumerate}

\section*{Acknowledgements}
This research was funded by the grant of Russian Science Foundation \textnumero21-72-00022, https://rscf.ru/en/project/21-72-00022/.\\

This research has made use of the Keck Observatory Archive (KOA), which is operated by the W. M. Keck Observatory and the NASA Exoplanet Science Institute (NExScI), under contract with the National Aeronautics and Space Administration.

We thank Vladimir Grinin for a discussion on the possible variability of the star's mid-IR flux. We also thank the anonymous referee for the thorough reading of the manuscript and the comments which helped to improve it.

\section*{DATA AVAILABILITY}

The data underlying this article will be shared on reasonable request to the corresponding author.

%%%%%%%%%%%%%%%%%%%%%%%%%%%%%%%%%%%%%%%%%%%%%%%%%%



\bibliographystyle{mnras}
\bibliography{references.bib} % if your bibtex file is called example.bib

%\begin{thebibliography}{99}
%\end{thebibliography}

%%%%%%%%%%%%%%%%%%%%%%%%%%%%%%%%%%%%%%%%%%%%%%%%%%

% Don't change these lines
\bsp	% typesetting comment
\label{lastpage}
\end{document}